\documentclass[%
 reprint,
 superscriptaddress,
nofootinbib,
 amsmath,amssymb,
 aps,
 prd,
 noeprint,
 floatfix
]{revtex4-2}

\usepackage{graphicx}% Include figure files
\usepackage{dcolumn}% Align table columns on decimal point
\usepackage{bm}% bold math
\usepackage{amssymb}
\usepackage{color}
\usepackage{siunitx}
\usepackage{subcaption}
\usepackage{hyperref}% add hypertext capabilities
\newcommand{\F}{\mathcal{F}}
\newcommand{\A}{\mathcal{A}}
\newcommand{\M}{\mathcal{M}}

\renewcommand{\S}{\mathcal{S}}
\newcommand{\dop}{\lambda}

\newcommand{\cosisq}{\cos^2\!\iota}
\renewcommand{\sc}[1]{\widehat{#1}}
\newcommand{\Fsc}{\sc{\F}}

\newcommand{\vrr}{\vec{r}}

\newcommand{\vn}{\hat{n}}

\newcommand{\seg}{_{\mathrm{seg}}}

\newcommand{\tref}{\tau_{\mathrm{ref}}}

\newcommand{\tSSB}{t_{_\mathrm{SSB}}}
\newcommand{\asini}{a_{\mathrm{p}}}
\newcommand{\argp}{\omega}
\newcommand{\tperi}{t_{\mathrm{p}}}
\newcommand{\tasc}{t_{\mathrm{asc}}}
\newcommand{\ecc}{e}
\newcommand{\Porb}{P_\mathrm{orb}}
\newcommand{\mmid}{_{\mathrm{m}}}
\newcommand{\tmid}{t\mmid}

\newcommand{\Nsft}{N_{\mathrm{SFTs}}}

\newcommand{\pf}{\textsc{pyfstat}}
\newcommand{\bb}{\textsc{bilby}}
\newcommand{\pt}{\textsc{ptemcee}}
\newcommand{\dn}{\textsc{dynesty}}
\newcommand{\ns}{\textsc{nessai}}
\newcommand{\ls}{\textsc{lalsuite}}
\newcommand{\sw}{\textsc{swig}}

\graphicspath{ {./figs/} {./} }

\usepackage{soul}

\begin{document}

\preprint{APS/123-QED}

\title{A new framework to follow up candidates from continuous gravitational-wave searches}

\author{P.~B.~Covas}
\affiliation{Max Planck Institute for Gravitational Physics (Albert Einstein Institute), D-30167 Hannover, Germany}
\affiliation{Leibniz Universität Hannover, D-30167 Hannover, Germany}

\author{R.~Prix}
\affiliation{Max Planck Institute for Gravitational Physics (Albert Einstein Institute), D-30167 Hannover, Germany}
\affiliation{Leibniz Universität Hannover, D-30167 Hannover, Germany}

\author{J.~Martins}
\affiliation{Max Planck Institute for Gravitational Physics (Albert Einstein Institute), D-30167 Hannover, Germany}
\affiliation{Leibniz Universität Hannover, D-30167 Hannover, Germany}

%\input{git_tag.tex}

%\date{\commitDATE; \commitIDshort-\commitSTATUS}
\date{\today}%

\begin{abstract}
  Searches for continuous gravitational waves from unknown neutron stars are limited in sensitivity due to their high computational cost. For this reason, developing new methods or improving existing ones can increase the probability of making a detection. In this paper we present a new framework that uses MCMC or nested sampling methods to follow-up candidates of continuous gravitational-wave searches. This framework aims to go beyond the capabilities of \pf{} (which is limited to the \pt{} sampler), by allowing a flexible choice of sampling algorithm (using \bb{} as a wrapper) and multi-dimensional correlated prior distributions. We show that MCMC and nested sampling methods can recover the maximum posterior point for much bigger parameter-space regions than previously thought (including for sources in binary systems), and we present tests that examine the capabilities of the new framework: a comparison between the \dn{}, \ns{}, and \pt{} samplers, the usage of correlated priors, and its improved computational cost.
\end{abstract}

\maketitle

\section{Introduction}
\label{sec:introduction}

Many searches for continuous gravitational waves (CWs) are routinely done, although there has been no detection yet (for a recent review of searches and methods see \cite{KeithReview}). All-sky searches for unknown neutron stars (both isolated and in binary systems) are the type of CW search with the highest computational cost, since all of the parameters describing the neutron star are unknown. For this reason, the sensitivity of these searches is the lowest since it is highly limited by the available computational budget \cite{WETTE2023102880}.

The main strategy followed by all-sky searches is to use a semi-coherent method (see \cite{Tenorio_2021} for a recent review). These methods separate the entire data-set into different segments between which phase continuity is not required, thus reducing the coherent time. Semi-coherent methods have been shown to be more sensitive than fully coherent methods at a limited computational budget \cite{PrixShaltev2011..optimalStackSlide}. After the main search is done with the initial set-up, candidates from the analyzed parameter space are marked as interesting, and several vetoes and a follow-up procedure are applied to them. The follow-up procedure increases the coherent time in a number of subsequent stages, checking at the end of each one if the candidate is consistent with the behavior expected from astrophysical CWs \cite{PhysRevD.61.082001,PhysRevD.72.042004,Shaltev_Prix_2013,PhysRevD.97.103020,Steltner_Papa_Eggenstein_Allen_Dergachev_Prix_Machenschalk_Walsh_Zhu_Kwang_2021}.

Most follow-up procedures work similarly to the main search, by setting a fixed template bank with a predetermined maximum loss of signal power, and calculating a detection statistic for each of these templates (see for example \cite{Steltner_Papa_Eggenstein_Allen_Dergachev_Prix_Machenschalk_Walsh_Zhu_Kwang_2021}). An alternative to this is to use a stochastic sampling algorithm, such as Markov chain Monte Carlo (MCMC) or nested sampling. This has already been applied to some CW searches, such as \cite{PhysRevLett.124.191102,PhysRevD.103.064017,Covas_2022,PhysRevD.89.124030}. The main advantage of these stochastic methods is that fewer evaluations of the detection statistic have to be carried out (if the parameter space is small enough for these methods to be able to localize the signal), thus lowering the computational cost of the follow-up. By lowering this cost, more candidates can be analyzed, thus increasing the sensitivity of a search. For this reason, improving and testing new follow-up methods is of utmost importance.

A stochastic sampling method applied to CWs searches has existed for a long time \cite{PhysRevD.70.022001,PhysRevD.72.102002}, and has been used in many targeted searches (such as \cite{LIGOScientific:2021hvc}), where all the parameters describing the phase evolution of the signal are known. This method only allows for fully coherent searches, thus not being feasible for the all-sky or directed searches that use semi-coherent detection statistics.

Until now, only one stochastic sampling framework for semi-coherent searches existed\footnote{Although a nested sampling algorithm has been used with a semi-coherent method for CWs \cite{SOAP}, our method is sensitive to much weaker signals.}, called \pf{} \cite{pyfstat}. The work presented in this paper improves and expands on \cite{PhysRevD.97.103020,pyfstat} in a number of ways. Firstly, we push the size of the parameter space to be analyzed to much higher values than previously reported \cite{PhysRevD.97.103020} (including signals from neutron stars in binary systems), showing that these algorithms can find the mode of the posterior in a reasonable time. Secondly, while previously a single MCMC sampler was used (\pt{} \cite{ptemcee}), now up to 8 nested sampling and 5 MCMC algorithms can be used and easily interchanged due to the flexibility of \bb{} \cite{bilby} (a Python package that we use as a link between the likelihood function and the sampler packages). Related to this, we introduce a new follow-up convergence criterion that can be used to compare the efficiency of different samplers. Thirdly, due to using \bb{}, there is also more flexibility in choosing a prior: \pf{} only allows priors set individually for each parameter, while \bb{} can use multi-dimensional correlated priors that can reduce the number of likelihood evaluations needed until convergence. Finally, we have also improved the computational efficiency of the underlying \ls{} \cite{lalsuite} likelihood implementation, thus further decreasing the amount of time needed for convergence.

Previous studies have focused on how to best set-up a chain of follow-up stages (from the initial stage to a fully coherent stage) \cite{PhysRevD.97.103020,PhysRevD.104.084012}, where the coherent time is increased at each stage. On the other hand, in this work we study the optimization of a single stage of the follow-up procedure. CW searches in a wide parameter space might generate a very large number of candidates, and the computational cost of the first stage of the follow-up procedure is usually much larger than the subsequent stages, thus constraining the sensitivity that can be achieved (by its dependence on the allowed number of candidates).

For this reason, decreasing the computational cost of the first stage is of critical importance. The computing cost of a single stage can be decreased by: 1) improving the choice of stochastic sampler and its tuning parameters; 2) reducing the size of the parameter space that needs to be searched; 3) instead of obtaining a precise posterior distribution, only finding a point close enough to the maximum posterior point (the size of the follow-up region for the next stage needs to be characterized with injections); 4) reducing the computational cost needed to calculate the likelihood function. In this paper we present tests that show how our new semi-coherent framework improves on these aspects.

This paper is organized as follows. In section \ref{sec:background} we introduce the signal model, the detection statistic, and its loss of signal power in searches with template banks. In section \ref{sec:framework} we present our new framework to perform Bayesian parameter estimation of semi-coherent CW searches, a new follow-up convergence criterion, and a model for its computational cost. In section \ref{sec:tests} we test our new framework by comparing different samplers and different prior distributions, and measuring the computational cost of each configuration. In section \ref{sec:conclusions} we present a summary and some ideas for future work.

\section{Background}
\label{sec:background}

\subsection{Signal model}

Rotating neutron stars with an asymmetry around their rotating axis emit CWs, which can be parametrized with four amplitude parameters $A$ and several phase-evolution parameters $\dop$. The four amplitude parameters $A$ consist of the overall signal amplitude $h_0$, the inclination angle $\iota$ between the line of sight and the neutron star rotation axis, the phase $\phi_0$ at a reference time $\tref$, and a polarization angle $\psi$. The phase-evolution parameters $\dop$ consist of the frequency of the signal $f$ (slowly changing over time as a function of a number of spin-down parameters), the sky position of the neutron star, and additional parameters describing its Newtonian orbital movement around the binary barycenter (3 parameters for a circular orbit and 5 for an elliptic orbit) if the neutron star is in a binary system \cite{2015arXiv150200914L}.

The CW signal depends non-linearly on the physical amplitude parameters $A$. In \cite{jks98:_data} a different set of four amplitude parameters $\A$ that linearize the functional form of the CW signal were found. This set of amplitude parameters allows one to write a CW signal $s^X(t)$ in the frame of detector $X$ as:
\begin{equation}
  \label{eq:signal}
  s^X(t;\A,\dop) = \sum_{\mu=1}^4 \A^{\mu}\, h^X_{\mu} (t; \dop),
\end{equation}
where $h^X_{\mu}$ are defined as:
\begin{equation}
  \label{eq:basisfunc}
  \begin{aligned}
    h_{1}^{X}&\equiv a^{X}(t; \dop) \, \cos \phi^X(t; \dop), \quad  h_{2}^{X} \equiv b^{X}(t; \dop) \,\cos \phi^X(t; \dop), \\
    h_{3}^{X}&\equiv a^{X}(t; \dop) \, \sin \phi^X(t; \dop), \quad  h_{4}^{X} \equiv b^{X}(t; \dop) \,\sin \phi^X(t; \dop),
  \end{aligned}
\end{equation}
in terms of the detector-frame signal phase $\phi^X(t)$ at time $t$, and the antenna-pattern functions $a^X(t)$ and $b^X(t)$ given in \cite{jks98:_data}.

The phase $\phi$ of a CW signal in the source frame can be expressed in terms of a Taylor expansion around a reference time $\tref$, namely:
\begin{equation}
  \phi(\tau) = \phi_0 + 2\pi \sum_{k=0}^{s} \frac{f_k}{(k+1)!} (\tau - \tref)^{k+1},
  \label{eq:phaseNS}
\end{equation}
where $\tau$ denotes the time in the source frame, and $s$ is the number of spin-down parameters $f_k$ needed to accurately describe the intrinsic frequency evolution, which are given by:
\begin{equation}
  f_k = \left. \frac{d^{k+1} \phi(\tau)}{d\tau^{k+1}} \right|_{\tau=\tref}.
  \label{eq:spindown}
\end{equation}
The phase can be transformed from the source frame to the detector frame by taking into account the movement of the neutron star and the movement of the detector with respect to the solar system barycenter (SSB). The barycentering relation between the different frames is obtained by first linking the wavefront-emission time $\tau$ in the source frame to its arrival time $\tSSB$ in the SSB frame, and then relating the SSB time to the arrival time $t$ at detector $X$ using the R{\o}mer-delay expression \cite{2015arXiv150200914L}:
\begin{align}
  \tau(\tSSB^X) &= \tSSB^X - R(\tau), \\
  \tSSB^X(t) &= t + \vrr^X(t)\cdot\vn,
  \label{eq:sourcedetector}
\end{align}
where $\vrr^X(t)$ is the position vector (in light-travel time) of detector $X$ with respect to the SSB, $\vn$ is the sky-position unit vector pointing from the SSB to the binary barycenter (BB), given by the right ascension $\alpha$ and declination $\delta$, and $R(\tau)$ is the radial distance (in light-travel time) from the source to the BB \cite{2015arXiv150200914L}, which depends on the five binary parameters $\asini$ (projected semi-major axis), $\Porb$ (orbital period), $\tasc$ (time of ascension) or $\tperi$ (time of periapsis), $\ecc$ (eccentricity), and $\argp$ (argument of periapsis).

\subsection{$\F$-statistic}
In order to detect CWs we distinguish between two hypotheses about the data $x(t)$: the data only consists of Gaussian noise $x(t)=n(t)$ ($\mathcal{H}_N$); the data consists of an astrophysical signal in addition to Gaussian noise $x(t) = n(t) + s(t;\A,\dop)$ ($\mathcal{H}_S$). The likelihood ratio $\mathcal{L}_r$ between these two hypothesis is \cite{cutler05:_gen_fstat}:
\begin{equation}
  \label{eq:loglikelihood}
  \begin{aligned}
    \ln \mathcal{L}_r (x; \A, \dop) &\equiv \A^{\mu} x_{\mu}-\frac{1}{2} \A^{\mu} \M_{\mu \nu} \A^{\nu}\,,
  \end{aligned}
\end{equation}
with implicit summation over $\mu,\nu=1,\ldots,4$, and we defined:
\begin{equation}
  \label{eq:xmu}
  x_{\mu} \equiv (x|h_\mu), \quad \M_{\mu\nu} \equiv (h_\mu|h_\nu)\,,
\end{equation}
in terms of the multi-detector scalar product \cite{cutler05:_gen_fstat,prix:_cfsv2} given by:
\begin{equation}
  \label{eq:4}
  (x|y) \equiv 2 \S^{-1} \sum_{X\alpha}^{\Nsft} \sqrt{w_{X\alpha}} \int\displaylimits_{t_{X\alpha}}^{t_{X\alpha} + T_{\mathrm{SFT}}} x_{X\alpha}(t)\, y_{X\alpha}(t)\, dt,
\end{equation}
where $\alpha$ is an index over the different short Fourier transforms\footnote{Several CW searches use SFTs \cite{Allen_Mendell} as the format for the input data. We assume stationary noise and constant antenna-pattern coefficients over the duration of each SFT.} (SFTs) per detector $X$ with duration $T_{\mathrm{SFT}}$. $\S(f)$ is the power spectral density, which represents the overall noise floor at frequency $f$ (the noise floor is assumed constant over the bandwidth occupied by the signal), defined as:
\begin{equation}
  \label{eq:54}
  \S^{-1} \equiv \frac{1}{\Nsft} \sum_{X\alpha} S_{X\alpha}^{-1},
\end{equation}
and $w_{X\alpha}$ is a per-SFT noise weight given by:
\begin{equation}
  \label{eq:5}
  w_{X\alpha} \equiv \frac{S_{X\alpha}^{-1}}{\S^{-1}}\,.
\end{equation}

The log-likelihood ratio depends quadratically on the amplitude parameters $\A^\mu$, and can thus be analytically maximized to yield the $\F$-statistic \cite{cutler05:_gen_fstat}:
\begin{equation}
    2 \F(x;\dop) \equiv 2 \, \max_{\A} \ln\mathcal{L}_r(x; \A, \dop) = x_{\mu}\,\M^{\mu\nu} x_{\nu}\,.
    \label{eq:2F}
\end{equation}
While the $\F$-statistic has underlying non-physical amplitude priors (and is thus not the optimal detection statistic) \cite{2009CQGra..26t4013P}, the analytical maximization results in a detection statistic with a lower computational cost compared to the physically motivated Bayesian counterpart. The $\F$-statistic has been used in many different searches for CWs, such as \cite{Covas_2022,Steltner_Papa_Eggenstein_Allen_Dergachev_Prix_Machenschalk_Walsh_Zhu_Kwang_2021}.

As explained in the introduction, due to the high computational cost of wide parameter-space CW searches the data to be analyzed is separated in different segments $N\seg$ of duration $T\seg$, and phase continuity is only required for the data within a segment. A semi-coherent likelihood can be obtained by taking the product between the likelihood in the different segments, assuming that they are independent \cite{PhysRevD.61.082001,prixetal2011:_transientCW}. Thus, we define the semi-coherent $\Fsc$-statistic as:
\begin{equation}
  \label{eq:10}
  2\Fsc \equiv \sum_{\ell=1}^{N\seg} 2\F_\ell\,,
\end{equation}
where $\F_\ell$ is the coherent $\F$-statistic computed on segment $\ell$.

The semi-coherent $\F$-statistic has a non-central $\chi^2$ distribution, with $4 N\seg$ degrees of freedom and a non-centrality parameter equal to the signal power $\rho^{2}$, which can be expressed as \cite{PhysRevD.98.084058}:
\begin{equation}
  \label{eq:rho2}
  \rho^{2} =  \frac{\Nsft T_{\mathrm{SFT}}}{\mathcal{D}^2} \,\left(\alpha_{1}\,A + \alpha_{2}\,B + 2 \alpha_{3}\,C \right),
\end{equation}
where we defined the sensitivity depth:
\begin{equation}
  \label{eq:sensdepth}
  \mathcal{D} = \frac{\sqrt{\S}}{h_0},
\end{equation}
the amplitude angle factors $\alpha_i(\iota,\psi)$:
\begin{equation}
  \label{eq:51}
  \begin{aligned}
    \alpha_{1} &\equiv \frac{1}{4}\left(1+\cosisq\right)^{2} \cos ^{2}2\psi+\cosisq\,\sin ^{2}2 \psi \\
    \alpha_{2} &\equiv \frac{1}{4}\left(1+\cosisq\right)^{2} \sin ^{2}2\psi+\cosisq\,\cos ^{2}2 \psi \\
    \alpha_{3} &\equiv \frac{1}{4}\left(1-\cosisq\right)^{2} \sin 2 \psi\, \cos 2 \psi,
  \end{aligned}
\end{equation}
and $A$, $B$, and $C$ are the components of the antenna-pattern matrix $\M_{\mu\nu}$.

Another important quantity is the critical ratio $\mathrm{CR}$ of the detection statistic (sometimes also called significance), which we define as \cite{Wette_2015}:
\begin{equation}
  \label{eq:SNR}
  \mathrm{CR} \equiv \frac{2\Fsc - \mu_N[2\Fsc]}{\sigma_N[2\Fsc]},
\end{equation}
where $\mu_N$ and $\sigma_N$ are the mean and standard deviation values of the distribution in the case of no signal present. When there is a signal, the expected value of the CR is:
\begin{align}
  \overline{\mathrm{CR}} &\equiv \frac{\mu_S[2\Fsc] - \mu_N[2\Fsc]}{\sigma_N[2\Fsc]} \nonumber \\
  &= \frac{4 N\seg + \rho^2 - 4 N\seg}{\sqrt{8 N\seg}} = \frac{\rho^2}{\sqrt{8 N\seg}},
  \label{eq:SNRExp}
\end{align}
a quantity that can characterize the dependence of the efficiency of a follow-up method on the signal power, in an independent way of the number of segments of the detection statistic. Another advantage of this quantity is that it is usually readily available for the candidates of a search (by using the estimated signal power defined in equation \eqref{eq:estpow}), while the sensitivity depth is not, because $h_0$ is not a directly available quantity from the search results.

\subsection{Mismatch and number of templates}
\label{sec:mismatch}

Due to the high computational cost of wide parameter-space CW searches, a grid of templates with finite spacing is used to cover the selected parameter-space region. For this reason, the values of the searched parameters will not be equal to the parameters of a possible astrophysical signal. The mismatch $\mu$ describes the relative loss of signal power $\rho^2$ due to not computing the detection statistic at the exact signal parameters $\lambda$:
\begin{align}
    \mu = 1 - \frac{\rho^2(\lambda')}{\rho^2(\lambda)},
    \label{eq:mismatch}
\end{align}
ranging from 0 (fully recovered signal power) to 1 (no recovered signal power), where $\lambda'$ represents the mismatched parameters. The mismatch decreases the recovered signal power, thus decreasing the sensitivity of a search.

The mismatch $\mu$ can be estimated with a Taylor expansion of the signal power around the signal parameters (where the mismatch attains a minimum of 0), keeping terms only up to second order \cite{prix06:_searc}: 
\begin{align}
    \mu \approx m \equiv g_{ij} (t; \lambda) d\lambda^i d\lambda^j + \mathcal{O}(d\lambda^3),
    \label{eq:phasemetric}
\end{align}
where $g_{ij}$ is a suitable parameter-space metric ($i$ and $j$ are indices over the different phase-evolution parameters). This equation represents a multi-dimensional ellipsoid with maximum mismatch given by $m$. This approximated mismatch can be bigger than 1, and it is known that it overestimates the true mismatch for mismatches $\mu$ higher than $\sim 0.3$ \cite{prix06:_searc,PhysRevD.88.123005}.

The phase metric in segment $\ell$ can be obtained by numerically integrating over time the derivatives of the phase model given by equation \eqref{eq:phaseNS} in the frame of the detector \cite{prix06:_searc}:
\begin{align}
  &g_{ij;\ell} (t; \lambda) = \nonumber \\ &\left < \frac{\partial \phi (t; \lambda)}{ \partial \lambda_i} \frac{\partial \phi (t; \lambda)}{ \partial \lambda_j} \right >_\ell - \left < \frac{\partial \phi (t; \lambda)}{ \partial \lambda_i} \right >_\ell \left<\frac{\partial \phi (t; \lambda)}{ \partial \lambda_j} \right >_\ell,
  \label{eq:metric}
\end{align}
where $\left < f(t) \right >_\ell \equiv 1 / T\seg \int_{t_\ell}^{t_\ell + T\seg} f(t) dt$. The semi-coherent metric can then be obtained by averaging the metrics $g_{ij;\ell}$ over the different segments \cite{Wette_2015}.

The number of templates $\mathcal{N}$ needed to cover a parameter-space region $\mathcal{R}$ of dimension $n$ with a maximum mismatch $m_0$ can be estimated with the following equation \cite{2007arXiv0707.0428P,PhysRevD.97.103020}:
\begin{align}
  \mathcal{N}(\theta,m_0) = \theta m_0^{-n/2} \int_{\mathcal{R}}  \sqrt{ \mathrm{det} (g_{ij}) } d^n \lambda,
  \label{eq:ntem}
\end{align}
where $\theta$ is the normalized thickness, a factor characterizing the lattice geometry. This equation assumes that the parameter-space region is much bigger than the spacing between templates, so that boundary effects can be neglected.

Further assuming that the metric components are constant, and that the maximum mismatch and normalized thickness are $m_0 = 1$ and $\theta = 1$, \cite{PhysRevD.97.103020} proposed an approximated quantity $\mathcal{N}^*$ to characterize the size of the parameter space:
\begin{align}
  \mathcal{N}^* \equiv \mathrm{Vol}(\mathcal{R}) \sqrt{ \mathrm{det} (g_{ij}) },
  \label{eq:nstar}
\end{align}
where $\mathrm{Vol}(\mathcal{R})$ is the coordinate volume of the parameter-space region.

In the next sections we will use $\mathcal{N}^*_{\mathrm{ell}}(m_{\mathcal{R}})$ when $\mathcal{R}$ is the metric ellipsoid described by equation \eqref{eq:phasemetric} with a maximum mismatch of $m_{\mathcal{R}}$, and $\mathcal{N}^*_{\mathrm{box}}$ when $\mathcal{R}$ is a box. Notice that two different quantities $m_0$ and $m_{\mathcal{R}}$ are used, where the former characterizes the spacing between the different templates, and the latter characterizes the size of a metric ellipsoid.

\section{Bayesian framework}
\label{sec:framework}

As explained in the introduction, we use the \bb{} package as an interface between the $\F$-statistic calculation and the different packages that carry out the stochastic sampling of the follow-up region. We choose \bb{} due to its ease of use and the large number of different samplers that it supports.

The posterior distribution for parameters $\dop$ is given by Bayes' theorem:
\begin{align}
  P(\dop | x, \mathcal{H}_S) = \frac{ \mathcal{L} (x | \dop, \mathcal{H}_S) P(\dop | \mathcal{H}_S) }{ P(x | \mathcal{H}_S) },
\end{align}
where $\mathcal{H}_S$ represents the signal hypothesis, $P(\dop | \mathcal{H}_S)$ is the prior of the phase-evolution parameters, $\mathcal{L}(x | \dop, \mathcal{H}_S)$ is the likelihood function for the signal hypothesis (after the amplitude parameters have been marginalized), and the constant factor in the denominator is sometimes called the evidence. We can divide this equation by the noise-only likelihood to obtain:
\begin{align}
  P(\dop | x, \mathcal{H}_S) \propto \frac{\mathcal{L} (x | \dop, \mathcal{H}_S)}{\mathcal{L} (x | \mathcal{H}_G)} P(\dop | \mathcal{H}_S).
\end{align}
It can be seen that the prior distribution and the likelihood ratio function are the two inputs that are needed for a parameter-estimation analysis, whose aim is the calculation of the posterior distribution.

\subsection{Likelihood}

As explained before, when limited by a certain computational budget a semi-coherent method can be more sensitive. In this case, we use the semi-coherent $\Fsc$-statistic (given by equation \eqref{eq:10}) as our underlying marginalized likelihood ratio function \cite{prixetal2011:_transientCW,PhysRevD.97.103020}:
\begin{align}
  \mathcal{L}_r (x | \dop) \equiv \frac{\mathcal{L} (x | \dop, \mathcal{H}_S)}{\mathcal{L} (x | \mathcal{H}_G)} \propto e^{\Fsc (x,\dop)}.
\end{align}
Other detection statistics can also be used as the likelihood function, such as one proposed to deal with non-Gaussian noise \cite{2014PhRvD..89f4023K}, a weighted $\F$-statistic \cite{Covas_2022}, or the short-segment dominant-response $\F$-statistic \cite{PhysRevD.105.124007}.

The shape of the likelihood as a function of the phase-evolution parameters is complicated by the presence of multiple correlations between these parameters, both local and global. While the local correlations are characterized by the Taylor-expanded phase metric given by equation \eqref{eq:phasemetric}, the global correlations are more difficult to characterize. An example of a global correlation is the relation between the frequency and the sky position, explored in detail in \cite{Prix_Itoh_2005}, where it can be seen that the likelihood has many local maxima in different regions of the parameter space, thus complicating the stochastic sampling procedure. Global correlations produced by the phase-evolution parameters describing the binary orbit of the neutron star have not yet been studied in detail but are also present. These correlations make the sampling of the posterior distribution more complicated, requiring the usage of samplers that are robust to posteriors with multiple modes.

We have written new \ls{} \cite{lalsuite} code that can efficiently calculate the semi-coherent $\Fsc$-statistic (and the other detection statistics mentioned above). We call this new semi-coherent log-likelihood function from \bb{} by using the \sw{} wrapper \cite{swig}.

\subsection{Prior}

Explicit priors for the phase-evolution parameters $\dop$ are needed. Since the parameter-estimation runs that we characterize in this paper are aimed as a follow-up from a previous stage of a search, we assume that we already have some information about these parameters and some account of their uncertainty, which defines the follow-up region $\mathcal{R}$. As discussed below, this region can be characterized by a multi-dimensional box or an ellipsoid, and we use two different distributions as our prior over $\mathcal{R}$: a uniform distribution or a Gaussian distribution.

\subsubsection{Follow-up region: box vs metric ellipse}
\label{sec:boxellipse}

In this section we compare the size of the follow-up region $\mathcal{R}$ when a box or a metric ellipsoid are used to describe it.

When carrying out a follow up of candidates from a search, the required size of the follow-up region can be obtained with injections, characterizing how far away are the candidates from these simulated signals. This can be done in two different ways: calculating the distances independently for each coordinate dimension $\lambda^i$, or calculating the mismatch given by equation \eqref{eq:phasemetric}. After this, the first method will set up a box, while the second method will instead set up a metric ellipsoid with size $m_{\mathcal{R}}$, both methods containing a desired percentage of the injections.

As an example, we test whether a box or an ellipse fits better (i.e. is smaller) a distribution of candidate offsets in $\{f_0,f_2\}$. To do this, we inject 1000 signals with random amplitude parameters at fixed $\rho^2=400$, random sky positions, and fixed $f_0$, $f_1$, and $f_2$, in simulated Gaussian noise. These signals are searched with a fine template grid of $m_0 = 5 \times 10^{-5}$ for the $\{f_0,f_2\}$ parameters, which have a strong correlation (we assume $\tref = \tmid$, where $\tmid$ is the mid-time of the observation span), and we obtain the grid point with the highest detection statistic. Then, we calculate the distance from this point to the signal, obtaining the required ellipsoid and box to contain all of the injections. 

\begin{figure}[htbp]
  \centering
  \includegraphics[width=\columnwidth]{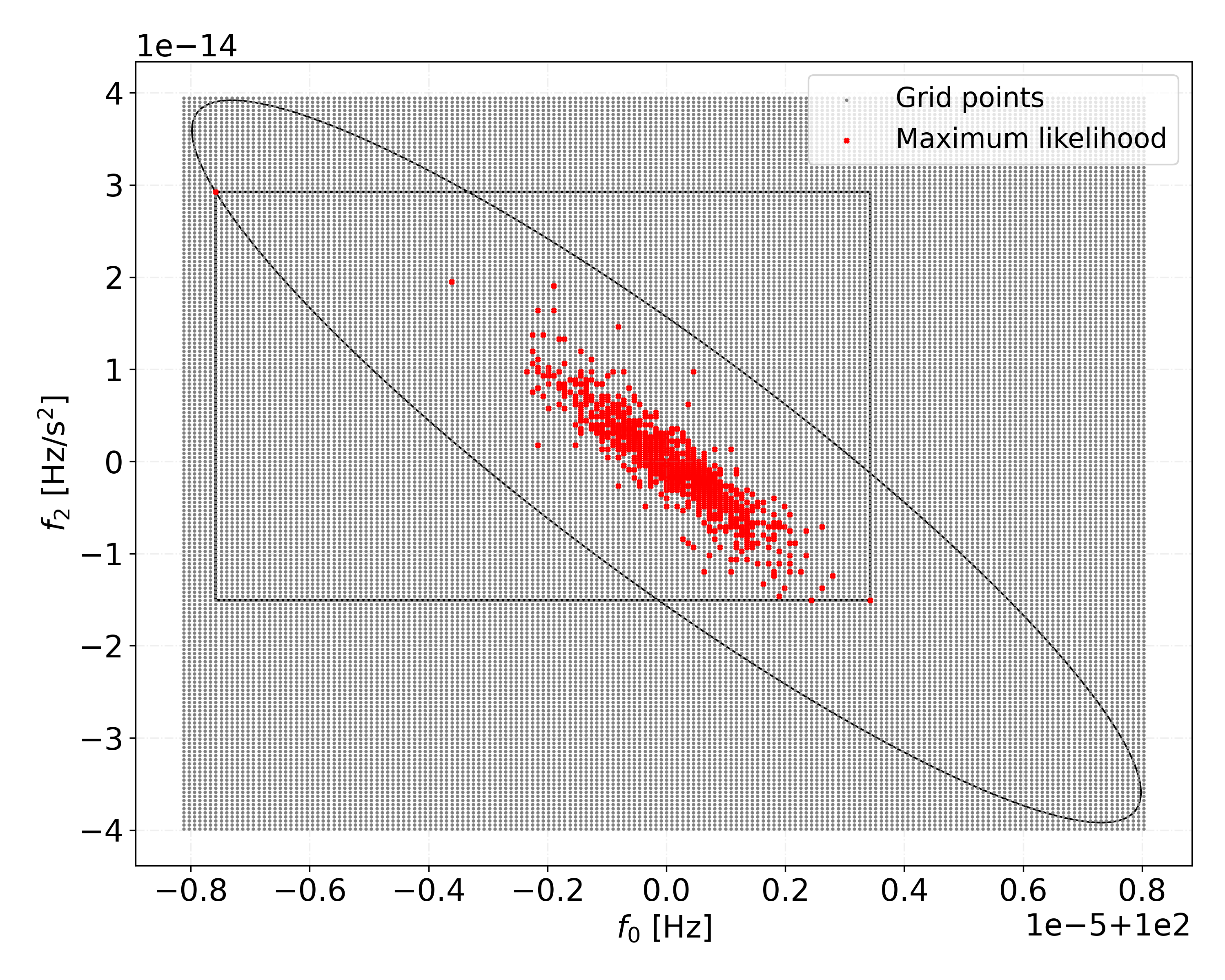}
  \caption{Plot showing the metric ellipsoid and the box required to contain 1000 injected signals, whose maximum likelihood point is shown with red markers (the injection is located at the center of the plot). The small gray circles show the rectangular grid of templates where we have calculated the detection statistic, with a maximum mismatch of $m_0 = 5 \times 10^{-5}$. This search has assumed $\tref = \tmid$.}
  \label{fig:sizesTemplated}
\end{figure}

The result of this test is shown in figure \ref{fig:sizesTemplated}. This figure indicates that using the ellipsoid results in a smaller number of templates, since $\mathcal{N}^*_{\mathrm{ell}} / \mathcal{N}^*_{\mathrm{box}} \simeq 0.8$. This result depends on the parameters that are searched: for instance, we observe that for the $\{f_0,f_1\}$ parameters using the ellipsoid results in a larger number of templates. The result might also depend on the percentage of injections that we want to cover with the follow-up region. We conclude that whether it is better to use a metric ellipsoid or a box depends on the search set-up and the specific parameter space that needs to be followed-up.

\subsection{Samplers}

In this work we test three different samplers: \pt{} \cite{ptemcee}, \dn{} \cite{dynesty}, and \ns{} \cite{nessai}. A general introduction to MCMC sampling algorithms can be found in \cite{Hogg_2018}, and a general introduction to nested sampling algorithms can be found in \cite{Ashton_2022}. Although \bb{} enables the usage of more samplers, here we focus on a direct comparison of these three samplers and on the optimization of their tuning parameters, leaving for future work the exploration of other stochastic samplers.

For all these samplers it is important to notice the difference between the termination criterion (the condition that determines when a parameter-estimation run is finished) and the convergence criterion (the condition that determines whether a parameter-estimation run has successfully characterized the posterior distribution). In this paper we propose a new follow-up convergence criterion (explained in section \ref{sec:convcrit}) that only takes into account the maximum posterior point (MP), instead of using a convergence criterion related to the posterior distribution.

An advantage of MCMC over nested sampling is that it is more straightforward to resume a run if the number of likelihood evaluations was not enough to reach convergence, since the walker chains can just be continued from their last positions in parameter space. For nested sampling, if the run has finished and the MP has been missed (e.g. the number of live points was too small), the run needs to be restarted, thus making the characterization of these methods more time-consuming.

\subsubsection{Ptemcee}
\label{sec:pte}

We use the \pt{} Python package \cite{ptemcee}, an MCMC sampler that has been used in previous CW follow-up searches \cite{PhysRevLett.124.191102,PhysRevD.103.064017,Covas_2022,PhysRevD.89.124030} (it is the only sampler available in \pf{}). We study the performance of \pt{} by changing the number of walkers $N_W$ and temperatures $N_T$, and for each of these configurations we find the total number of steps $N_{\mathrm{MCMC}}$ that is required so that all injections have reached our follow-up convergence criterion (defined in section \ref{sec:convcrit}). We have modified the \pt{} code in \bb{} so that no estimation of the auto-correlation length is done, in order to decrease the number of likelihood evaluations, thus imitating the behaviour of the \pf{} package. The termination criterion for a single run is the number of steps $N_{\mathrm{MCMC}}$.

\subsubsection{Dynesty}
\label{sec:dyn}

We use the \dn{} Python package \cite{dynesty}, a nested sampler that has been used for CW targeted searches and which is broadly used within the gravitational-wave community. We study the performance of \dn{} by changing the number of live points and the point-finding algorithm (called sample method in the \bb{} documentation): we use the \textit{rslice} point-finder, the \textit{act-walk} point-finder, and a custom point-finder that only proposes new points until a point with a likelihood higher than the current threshold is found. For each of these point-finding algorithms we change some of their tuning parameters such as the number of slices or the number of auto-correlation lengths. We use the default value for the termination criterion, which is that the estimated remaining log-evidence $\Delta \ln{P(x | \mathcal{H}_S)}$ is 0.1, and we find the number of live points that is required so that all injections
have reached our follow-up convergence criterion (defined in section \ref{sec:convcrit}). We have modified the \dn{} code so that no resampling is done at the end of the run, since we are indifferent to the statistical independence of the obtained samples, but do not want to lose the sample with the highest likelihood.

\subsubsection{Nessai}
\label{sec:nessai}

We use the \ns{} Python package \cite{nessai}, a nested sampler that makes use of normalizing flows and has shown more efficiency than \dn{} for some problems. We study the performance of \ns{} by changing the number of live points. We use the default value for the termination criterion, which is that the estimated remaining log-evidence $\Delta \ln{P(x | \mathcal{H}_S)}$ is 0.1, and we find the number of live points that is required so that all injections have reached our follow-up convergence criterion (defined in section \ref{sec:convcrit}). We have modified the \ns{} code so that no resampling is done at the end of the run, since we are indifferent to the statistical independence of the obtained samples, but do not want to lose the sample with the highest likelihood.

\subsection{Follow-up convergence criterion}
\label{sec:convcrit}

Although these samplers can be used to obtain precise posterior distributions, we simply aim to find a point close enough to the maximum posterior point (MP). This is because at the first follow-up stage we are not interested in describing the posterior distribution of the phase-evolution parameters with high precision, but only that the MP can be used to check whether the candidate behaves as an astrophysical signal and if so, to be used as an input to the subsequent follow-up stage. A disadvantage of this strategy is that since the posterior has not necessarily converged, it cannot be used as the prior for the subsequent stage of the follow-up. For this reason, injections are required to characterize the uncertainty of the MP if more than one follow-up stages are needed.

In a previous CW follow-up study \cite{PhysRevD.97.103020} a convergence criterion (related to the Gelman-Rubin statistic) for the posterior was used. In this work, instead, we propose a convergence criterion not related to the posterior, but only to the MP. Follow-up convergence is considered to be achieved when the estimated signal power of the candidate $\hat{\rho}^2_{\mathrm{cand}}$ is close enough to the estimated signal power of the MP $\hat{\rho}^2_{\mathrm{MP}}$. The estimated signal power $\hat{\rho}^2$ is defined as:
\begin{align}
  \hat{\rho}^2 \equiv 2\Fsc - \mu_N[2\Fsc] = 2\Fsc - 4 N\seg.
  \label{eq:estpow}
\end{align}
In practice it can be difficult to find the MP even for an injection, because of noise shifting the signal peak. We approximate $\hat{\rho}^2_{\mathrm{MP}}$ for each injection by calculating the $2\Fsc$ value at the signal parameters, i.e. $\hat{\rho}^2_{\mathrm{MP}} \sim \hat{\rho}^2_{\mathrm{inj}}$, which turns out to be a good approximation.

The convergence criterion is defined as:
\begin{align}
  c \equiv 2\frac{\hat{\rho}^2_{\mathrm{cand}} - \hat{\rho}^2_{\mathrm{MP}}}{\hat{\rho}^2_{\mathrm{cand}} + \hat{\rho}^2_{\mathrm{MP}}} > c_0.
  \label{eq:convergence}
\end{align}

An advantage of this follow-up convergence criterion is that it makes a comparison of efficiency between different samplers straightforward. This criterion can be used in a real search by first characterizing the tuning parameters of the samplers that are required so that a certain fraction of injections are converged. This is the procedure we follow in the tests done in sections \ref{sec:nstar}, \ref{sec:comparing}, and \ref{sec:priorComparison}, requiring that all injections have converged.

\begin{figure}[htbp]
  \centering
  \includegraphics[width=\columnwidth]{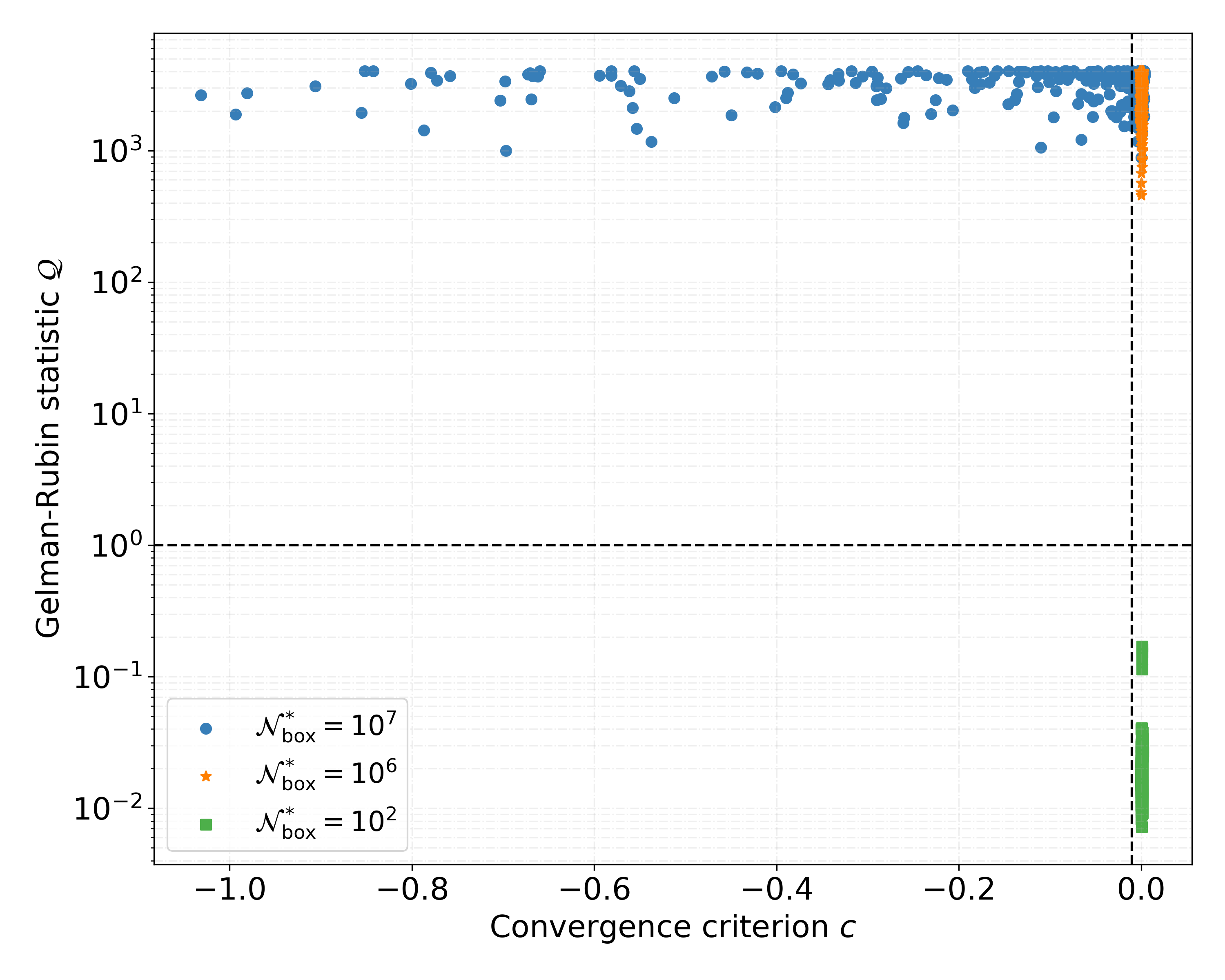}
  \caption{Follow-up convergence criterion $c$ given by equation \eqref{eq:convergence} and correspondent Gelman-Rubin statistic $\mathcal{Q}$ given by equation 28 in \cite{PhysRevD.97.103020}. Each point is the result of a \pt{} run over the $\{f_0,f_1\}$ parameters, with 100 walkers, 3 temperatures, and 4000 steps. The different markers show different sizes of the parameter space. The vertical dashed line shows the threshold at $c_0 = -0.01$, while the horizontal dashed line shows the threshold at $\mathcal{Q} = 1$ from \cite{PhysRevD.97.103020}.}
  \label{fig:convergence} 
\end{figure}

Figure \ref{fig:convergence} shows a plot comparing the Gelman-Rubin statistic $\mathcal{Q}$ (given by equation 28 of \cite{PhysRevD.97.103020}, which compares the between-walker variance and the within-walker variance) and our convergence criterion for the \pt{} sampler, for the tests explained in section \ref{sec:nstar}. We can clearly see when a run has not found the MP, as happens with $\mathcal{N}^*_{\mathrm{box}} = 10^7$, indicating that the number of likelihood evaluations was not enough. We can also see a parameter-space size with $\mathcal{N}^*_{\mathrm{box}} = 10^6$ that the $\mathcal{Q}$ criterion would classify as non-converged (i.e. much higher than 1) but our follow-up criterion classifies as converged. We remark again that these criteria refer to different aspects: $\mathcal{Q}$ is a posterior convergence criterion, while we have defined a follow-up convergence criterion that only checks for the MP. The advantage of our criterion comes from the fact that fewer likelihood evaluations are required to find the MP than to properly characterize the posterior.

\subsection{Computational model}
\label{sec:compmodel}

In this section we present a model for the computational cost of a single follow-up stage. We divide the cost per likelihood evaluation $\tau_i$ into two contributions, following the description of \cite{PhysRevD.97.103020}: the time to compute the detection statistic $\tau_{\mathrm{F}}$, and the remaining overhead time $\tau_{\mathrm{O}}$ (due to various factors such as proposing a new point in parameter space and other sampler-specific computations):
\begin{align}
 \tau_i= \tau_{\mathrm{F}} + \tau_{\mathrm{O}}.
 \label{eq:taui}
\end{align}

The overhead time $\tau_{\mathrm{O}}$ depends on the sampling method and on the prior distributions. If the detection statistic is $2\Fsc$, the timing model is given by\footnote{We use the \textit{Demod} implementation of the $\F$-statistic \cite{Williams:1999nt,prix:_cfsv2}.} \cite{prix:Fstat_timing}:
\begin{align}
  \label{eq:detstat}
  \tau_{\mathrm{F}} &= \tau_{\mathrm{C}} + \tau_{\mathrm{B}} \\
  \tau_{\mathrm{B}} &= \tau_{\mathrm{Antenna}} + \tau_{\mathrm{Sky}} + \tau_{\mathrm{Binary}},
  \label{eq:buffer}
\end{align}
where $\tau_{\mathrm{C}}$ is the time to compute the core $\F$-statistic quantities and $\tau_{\mathrm{B}}$ is the buffer time needed to compute auxiliary quantities such as the antenna-pattern coefficients $\tau_{\mathrm{Antenna}}$, and the source-to-SSB $\tau_{\mathrm{Binary}}$ and SSB-to-detector $\tau_{\mathrm{Sky}}$ barycentering transformations. Both $\tau_{\mathrm{F}}$ and $\tau_{\mathrm{B}}$ depend linearly on the number of SFTs $N_{\mathrm{SFTs}}$.

The total time to complete a follow-up stage for a single candidate is:
\begin{align}
   \tau = \sum_{i=1}^{N_{\mathcal{L}}} \tau_i = N_{\mathcal{L}} \bar{\tau},
    \label{eq:tau}
\end{align}
where $\bar{\tau} = \left< \tau_i \right>$ is the average over the different likelihood calculations and $N_{\mathcal{L}}$ is the number of likelihood evaluations, which depends on the size of the parameter space, the critical ratio $\overline{\mathrm{CR}}$, and the efficiency of the sampling algorithm. With this model we are assuming that other contributions to the total timing (such as I/O) are negligible. In section \ref{sec:comptests} we will test and evaluate this timing model.

\section{Testing the new framework}
\label{sec:tests}

In this section we present several tests that illustrate the performance and capabilities of the new framework. The follow-up regions $\mathcal{R}$ of these tests are given by bounding boxes of metric ellipsoids with different $m_{\mathcal{R}}$ values. This is done so that all injections have the same $\mathcal{N}^*_{\mathrm{box}}$ and to ensure that all parameters are fully resolved.

\subsection{Maximum size of follow-up region}
\label{sec:nstar}

In this section we show that stochastic sampling methods are able to find the MP for much larger parameter spaces than previously demonstrated. In \cite{PhysRevD.97.103020} it was discussed that the maximum parameter-space size for proper posterior convergence was around $\mathcal{N}^*_{\mathrm{box}} \sim 10^3$ (for the sampler configurations that were investigated), showing a specific example with $\mathcal{N}^*_{\mathrm{box}} = 10^7$ where the Gelman-Rubin criterion indicated non-convergence.

In order to illustrate that the MP can be recovered in much larger parameter-space regions, we use again the \pt{} sampler with 100 walkers and 3 temperatures, a configuration that was used in \cite{PhysRevD.97.103020}. We simulate a single CW signal with 100 different noise realizations, and run the algorithm 5 times for each of these noise realizations with a different sampler random seed, to take into account the intrinsic randomness of the stochastic sampler. We do this for different sizes of the follow-up region (from $\mathcal{N}^*_{\mathrm{box}}=10^2$ to $\mathcal{N}^*_{\mathrm{box}}=10^8$), searching a box in the $\{f_0,f_1\}$ parameters. We use the new convergence criterion proposed in section \ref{sec:convcrit}, requiring that all the 500 runs have a follow-up convergence value $c$ higher than $c_0=-0.01$ (i.e. we take as many steps as necessary to reach convergence). We simulate Gaussian noise with a duration of 10 days from a single detector.

\begin{figure}[htbp]
  \centering
  \includegraphics[width=\columnwidth]{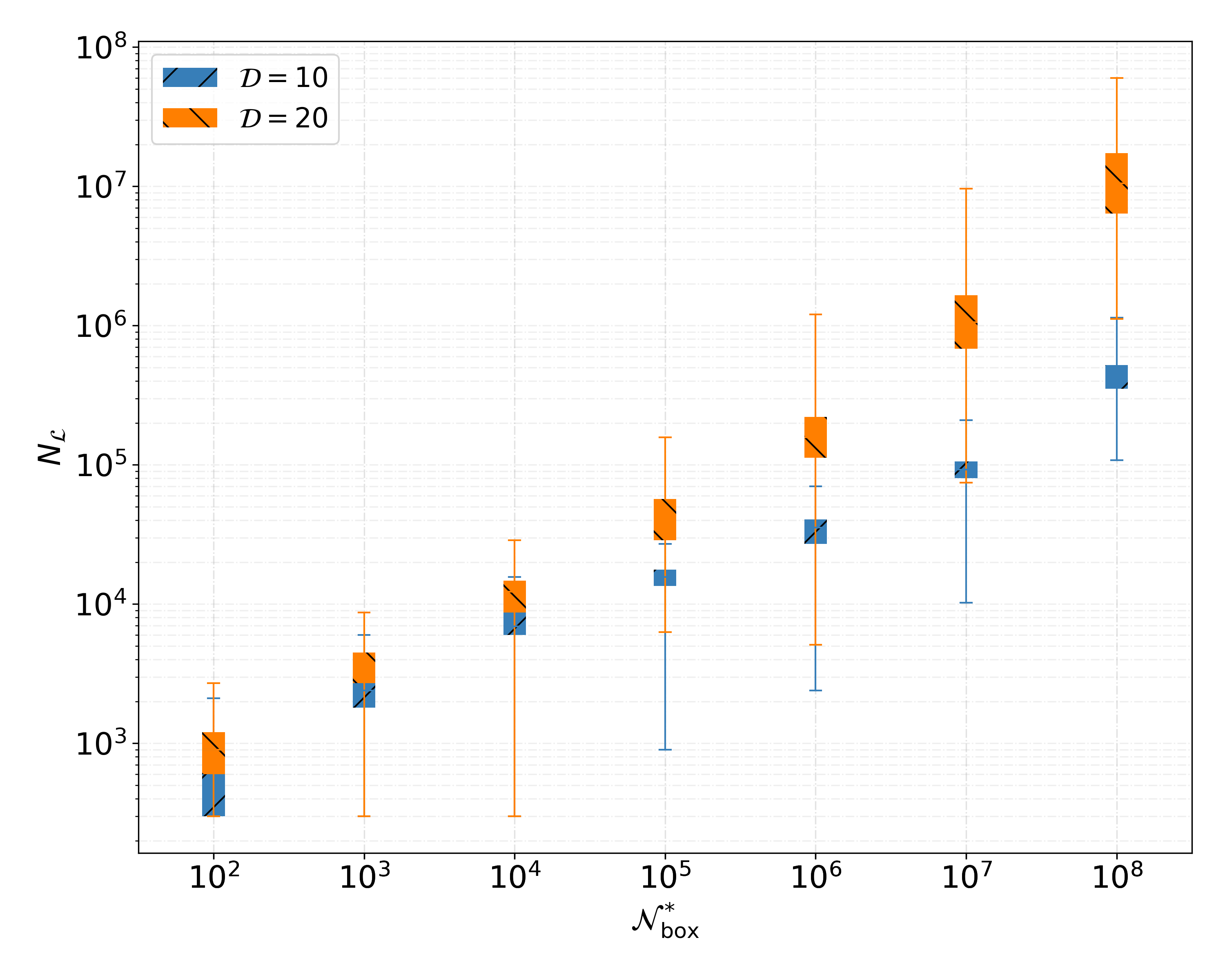}
  \caption{Number of likelihood evaluations $N_{\mathcal{L}}$ as a function of the number of unit-mismatch templates $\mathcal{N}^*_{\mathrm{box}}$ given by equation \eqref{eq:nstar} for the \pt{} sampler with 100 walkers and 3 temperatures, for a search over the $\{f_0,f_1\}$  parameters. Each color shows a different value of the sensitivity depth $\mathcal{D}$. We have used the convergence criterion $c>-0.01$. Each box extends from the 25 to the 75 quartiles, while the vertical lines go to the minimum and maximum of each distribution.}
  \label{fig:NlikeNstar}
\end{figure}

The results are shown in figure \ref{fig:NlikeNstar}, where the required number of likelihood evaluations can be seen as a function of $\mathcal{N}^*_{\mathrm{box}}$, for two different signal strengths. It can be seen that the weaker signal required more likelihood evaluations. We also observe that as $\mathcal{N}^*_{\mathrm{box}}$ increases, the ratio of $N_{\mathcal{L}}$ between the two different signal strengths increases as well. This figure shows that stochastic samplers can deal with parameter-space regions much bigger than previously found in \cite{PhysRevD.97.103020}, mainly due to the usage of a different convergence criterion (requiring convergence of the MP and not of the posterior) and to a higher number of likelihood evaluations (in \cite{PhysRevD.97.103020}, the number of likelihood evaluations was limited to $N_{\mathcal{L}} = 180000$, while here we have taken as many steps as necessary to reach convergence).

These results allow us to reconsider the optimal procedure when designing a follow-up with multiple stages and a limited computational budget. Recent works have built a ladder of coherent times \cite{PhysRevD.97.103020,PhysRevD.104.084012}, where the number of follow-up stages is minimized by using the maximum possible increase in coherent time while constraining the size of the follow-up region to be less than the maximum allowed by some posterior convergence criterion. As we have observed, however, there is no hard limit on the size of the follow-up region, and the required number of likelihood evaluations also depends on the strength of the signal compared to noise (characterized by $\overline{\mathrm{CR}}$, which will increase in subsequent stages of the follow-up). Because of this, the optimal construction of the ladder of coherent times given a limited computational budget will likely follow a different procedure than the one proposed by these previous studies. We leave a more detailed investigation of this issue for future work.

\subsection{Comparing samplers}
\label{sec:comparing}

In this section we compare the performance of the \pt{}, \dn{}, and \ns{} samplers as a function of some of their tuning parameters. Both nested and MCMC samplers have many tuning parameters that govern the behavior of the algorithm, such as the number of live points or the number of parallel walkers. Our tests investigate the configuration that results in the best compromise between run-time and recovery of injections. The best algorithm will be the one for which all injections reach our follow-up convergence criterion with the lowest number of likelihood evaluations (assuming that the time spent outside of the likelihood function is negligible with respect to the total time, which as shown in section \ref{sec:comptests} is the case for a realistic amount of data).

\subsubsection{Set-up}

We perform four different tests with parameters shown in table \ref{tab:pars}. The number of unit-mismatch templates $\mathcal{N}^*_{\mathrm{box}}$ for each test is given in the leftmost column of that table. We use a uniform prior for each parameter. T1 and T2 exemplify a smaller parameter-space region but with more complicated correlations due to the $\alpha$ and $\delta$ coordinates, while T3 and T4 have smaller correlations between their parameters. T2 uses signals with smaller amplitude, while T4 has a higher number of unit-mismatch templates. In these tests we have used 100 different signals, and run the algorithm 5 times for each of these signals with a different sampler random seed, to take into account the intrinsic randomness of the stochastic sampler. We have used a single segment of coherent time $T\seg = 864000$ s, but the results shown here are independent of the number of segments (if the expected critical ratio $\overline{\mathrm{CR}}$ of the signals is scaled accordingly). For the tests in this section, we use a convergence threshold of $c_0=0$.

\begin{table}[htbp]
\begin{tabular}{c | c c c c }
            & $\mathcal{N}^*_{\mathrm{box}}$ & $\left< \mathcal{N}^*_{\mathrm{ell}} \right>$  & Parameters                                     & $\rho^2$ \\ \hline
\textbf{T1} & $10^6$                         & 4                                              & $\{f_0,f_1,\alpha,\delta\}$                    & 85       \\ 
\textbf{T2} & $10^6$                         & 4                                              & $\{f_0,f_1,\alpha,\delta\}$                    & 57       \\
\textbf{T3} & $10^6$                         & 1                                              & $\{f_0,f_1,\asini,\Porb,\tasc,\ecc,\argp\}$    & 85       \\
\textbf{T4} & $10^9$                         & 143                                            & $\{f_0,f_1,\asini,\Porb,\tasc,\ecc,\argp\}$    & 85       \\
\end{tabular}
\caption{Parameters of the different tests performed in sections \ref{sec:comparing} and \ref{sec:priorComparison}. All sets generate fake Gaussian noise for $N_{\mathrm{SFTs}} = 480$ with one detector, with a single segment of duration 10 days. The injections have random amplitude parameters, with an amplitude $h_0$ given so that their signal power is equal to $\rho^2$, and are isotropically distributed over the sky, with fixed $f_0=100$ Hz, $f_1 = -10^{-11}$ Hz/s, $f_2 = 10^{-23}$ Hz/s$^2$, $\asini=10$ l-s, $\Porb = 10$ days, $\ecc=0.3$, $\argp=2$, $\tasc = \tmid - 275019$, and $\tref = \tmid$, where $\tmid$ is the mid-time of the observation span. $\mathcal{N}^*_{\mathrm{box}}$ is the number of unit-mismatch templates contained in the follow-up regions $\mathcal{R}$ used in section \ref{sec:comparing}, while $\left< \mathcal{N}^*_{\mathrm{ell}} \right>$ is the (averaged over the different injections) number of unit-mismatch templates contained in the follow-up regions used in section \ref{sec:priorComparison}.}
\label{tab:pars}
\end{table}

For \pt{}, we test 9 different combinations of tuning parameters: $N_W = 20,100,1000$ and $N_T = 1,3,6$, for which the temperatures are logarithmically spaced between 1 and $10$ for the first two temperature values and between 1 and 100 for the last one. For \dn{}, we test different numbers of live points: we search for the injections, and if they all converge, we divide the number of live points by 1.5, otherwise we multiply the number of live points by 1.5. In this way, we try to find the optimal number of live points within a factor of 1.5. For \ns{}, we use the default tuning parameters, and use the same strategy as with \dn{} to find an approximately optimal number of live points.

\subsubsection{Results}

In this section we present the results of our tests. For \pt{}, these indicate that the best configuration for all the test sets is the one with $N_T = 1$ and $N_W = 100$. For \dn{}, the tests indicate that the best configuration is the \textit{act-walk} point-finder method, after reducing the tuning parameter \textit{nact} to 1 and the maximum number of walks \textit{maxmcmc} to 100. The number of required live points so that all injections reached our follow-up convergence criterion was: for T1, 150; for T2, 200; for T3, 100; for T4, 150. For \ns{}, the number of required live points was: for T1, 150; for T2, 300; for T3, 300; for T4, 900.

\begin{figure}[htbp]
  \centering
  \includegraphics[width=\columnwidth]{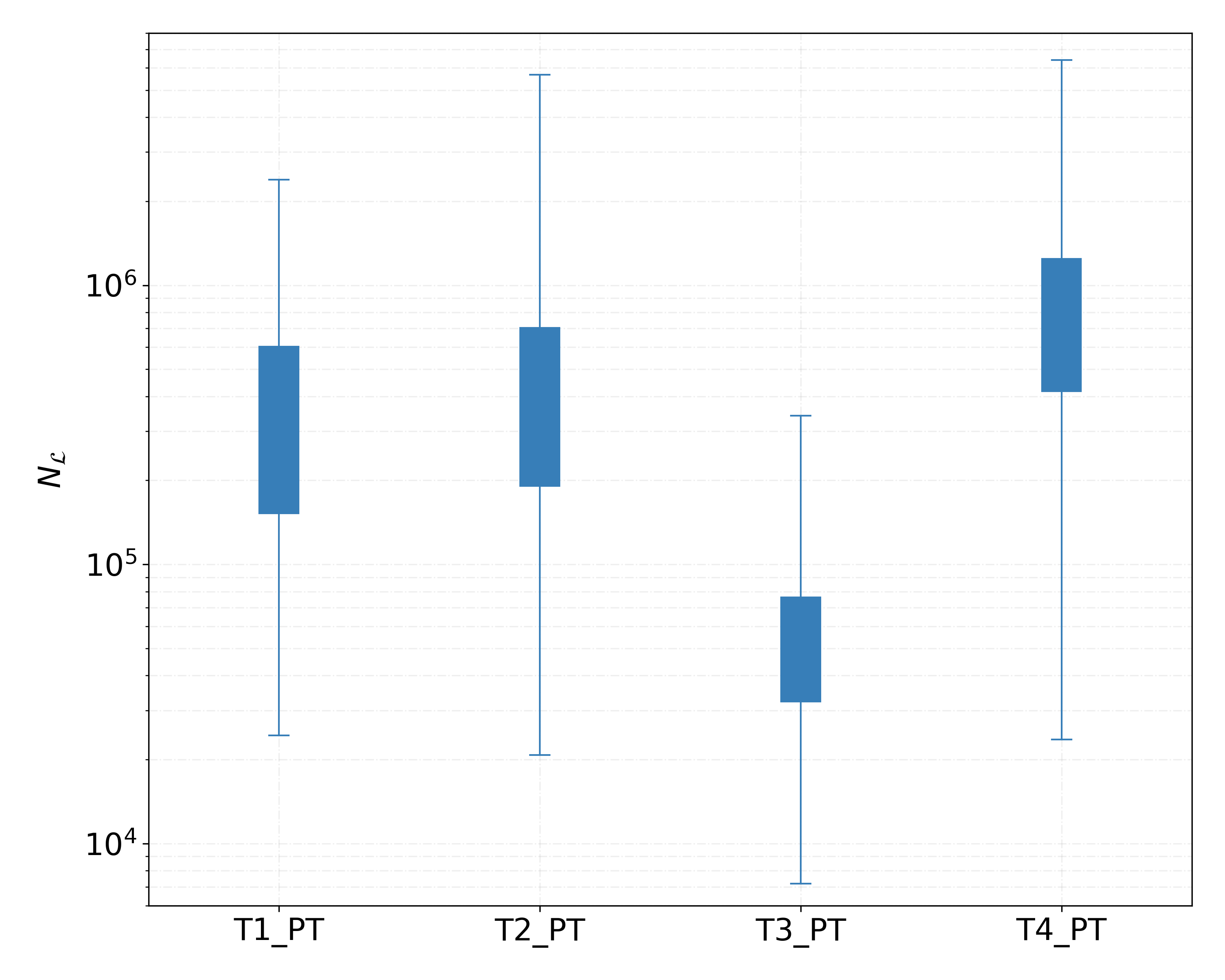}
  \caption{Number of likelihood evaluations required to reach follow-up convergence as a function of the test set for the best \pt{} configuration. Each box extends from the 25 to the 75 quartiles, while the vertical lines go to the minimum and maximum of each distribution.}
  \label{fig:rescomPt}
\end{figure}

\begin{figure}[htbp]
  \centering
  \includegraphics[width=\columnwidth]{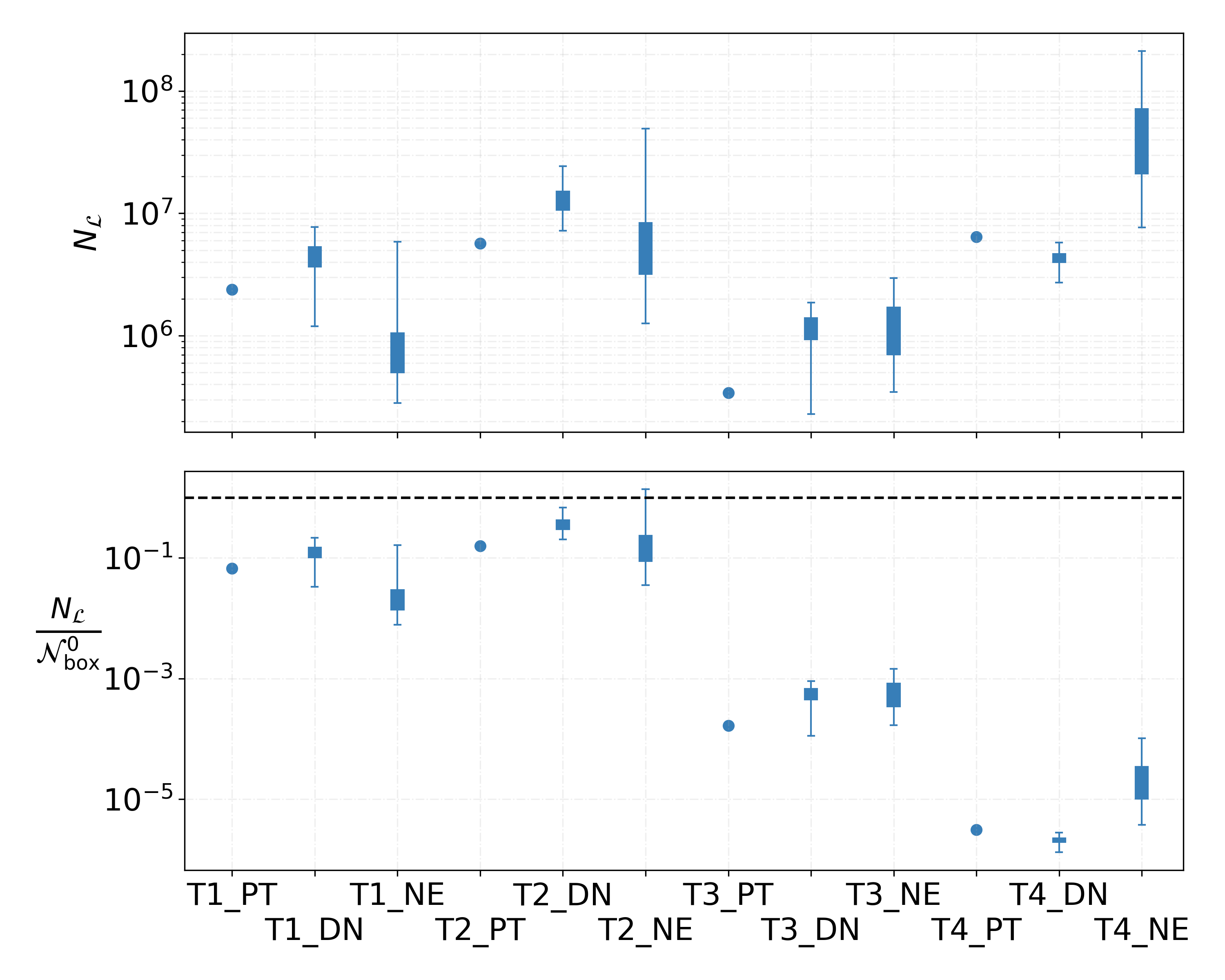}
  \caption{The upper plot shows the number of likelihood evaluations as a function of the sampler and test set, for the best sampler configurations. The labels on the x-axis refer to the tests of table \ref{tab:pars} and the different samplers, where PT is \pt{}, NE is \ns{}, and DN is \dn{}. The lower plot shows the same quantity but divided by $\mathcal{N}^{\,0}_{\mathrm{box}} \equiv \mathcal{N}(A^*_n,m_0=0.1)$, which is the number of templates required by a deterministic template bank with a maximum mismatch $m_0 = 0.1$ and an $A_n^*$ lattice. Each box extends from the 25 to the 75 quartiles, while the vertical lines go to the minimum and maximum of each distribution.}
  \label{fig:rescomT}
\end{figure}

For \pt{}, figure \ref{fig:rescomPt} shows the results for all the injections, instead of only the highest $N_{\mathcal{L}}$ (which is used to compare the efficiency to \dn{} and \ns{}). Using the highest number of likelihood evaluations as the termination criterion for \pt{} is required to recover all injections, and that is why only the highest $N_{\mathcal{L}}$ is shown in figure \ref{fig:rescomT}.

\begin{figure}[htbp]
  \centering
  \includegraphics[width=\columnwidth]{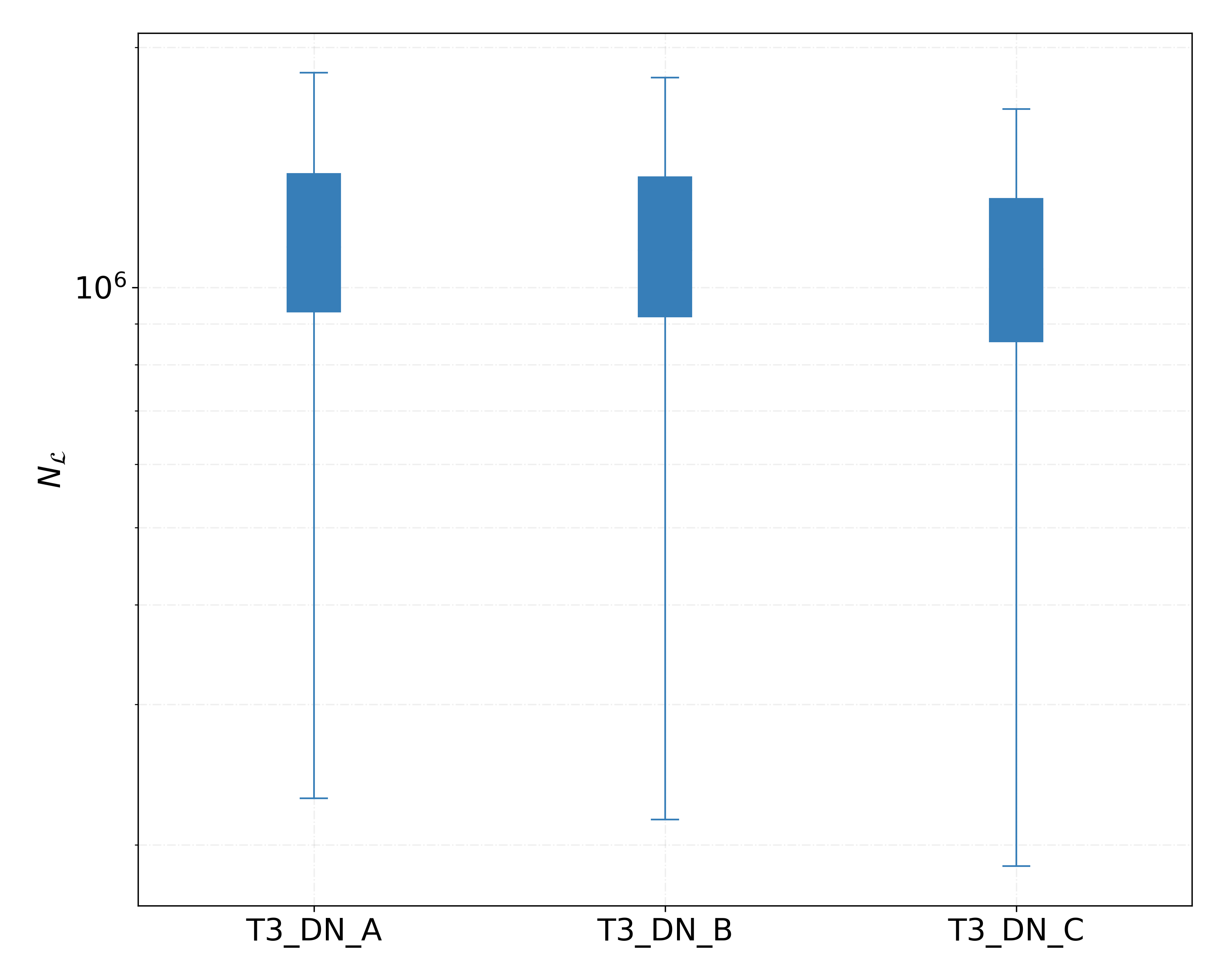}
  \caption{Number of likelihood evaluations required to reach the termination criterion for T3 as a function of different \dn{} termination criteria, with an equal number of live points. Method A uses the default $\Delta \ln{P(x | \mathcal{H}_S)} = 0.1$, method B uses $\Delta \ln{P(x | \mathcal{H}_S)} = 1.0$, and method C uses the number of likelihood evaluations needed to reach $c > 0.0$ for each injection, as with \pt{}. Each box extends from the 25 to the 75 quartiles, while the vertical lines go to the minimum and maximum of each distribution.}
  \label{fig:rescomDN}
\end{figure}

\dn{} can also use a certain number of likelihood evaluations as its termination criterion (and then the highest $N_{\mathcal{L}}$ across all injections would need to be used), but we notice that this reduces the efficiency, as shown with the comparison between the leftmost and rightmost distributions in figure \ref{fig:rescomDN}, which presents a comparison between different \dn{} termination criteria. This figure also shows that using a remaining log-evidence of 1 instead of 0.1 almost does not modify the number of likelihood evaluations. Using a number higher than 1 results in some injections not reaching follow-up convergence, thus requiring a higher number of live points, which decreases the efficiency.

The number of likelihood evaluations $N_{\mathcal{L}}$ obtained with the best configurations of the \pt{}, \dn{}, and \ns{} samplers at the four test sets is shown in the upper plot of figure \ref{fig:rescomT}. For \pt{}, these are the number of likelihood evaluations required to reach our follow-up convergence criterion for all injections (i.e. we run \pt{} until convergence), while for \dn{} and \ns{} these are the number of likelihood evaluations obtained after the default termination criterion is met (where all injections have reached the follow-up convergence criterion).

The results presented in figure \ref{fig:rescomT} show some interesting features: for T1 and T2 \ns{} performed better, while for T4 it performed much worse than the other samplers. This indicates that \ns{} has a worse performance when the signal occupies a small fraction of the follow-up region, since training the normalizing flows is not effective due to most live points sampling regions of parameter space where only noise is present. For T4 \dn{} required fewer likelihood evaluations than the other samplers, which indicates that \dn{} is more robust than the other samplers to the decrease of the fraction of follow-up region occupied by the signal.

The results clearly show that when the amplitude of the signals is smaller (from T1 to T2), more likelihood evaluations are needed for all samplers, as already shown in the results of section \ref{sec:nstar}. The results indicate that \pt{} is less affected than \dn{} and \ns{} when the signal strength is reduced. When the size of the parameter space is increased (from T3 to T4), the results also show an increase in the number of likelihood evaluations, as expected. On the other hand, when the size of the parameter space is increased but the coordinates have different correlations (from T1 to T4), the \dn{} results do not clearly show an increase in the number of likelihood evaluations. These results show again that the size of the parameter space is not the only property that characterizes the required number of likelihood evaluations, but also the searched parameters (due to their influence on the shape of the likelihood function) and the critical ratio $\overline{\mathrm{CR}}$ of the signal.

We remark that if not all the injections are required to reach follow-up convergence (as in many searches where a small false dismissal is allowed), the number of likelihood evaluations that is needed would be lower for all samplers. For \pt{}, this is because we use the highest $N_{\mathcal{L}}$, which would be lower if a fraction of the injections are allowed to not reach convergence. For \dn{} and \ns{}, this is because a smaller number of live points could be used, and thus the average $N_{\mathcal{L}}$ would decrease. 

We have repeated the tests but shifting the center of the prior from the signal parameters, in order to test the safeness of the previous signal-centered tests. We have seen that the results are comparable, thus indicating that centering the priors on the signal parameters does not introduce any noticeable bias.

In the lower plot of figure \ref{fig:rescomT} we show a comparison between the required number of likelihood evaluations for the stochastic samplers and the number of templates that would be needed in a search with a template bank with a maximum mismatch $m_0 = 0.1$ and an $A_n^*$ lattice. It can be seen that for the two first test sets the decrease in the number of required likelihood evaluations by using stochastic samplers instead of a deterministic template bank is around one order of magnitude or less. When more dimensions are included (as shown by T3 and T4) this ratio is around three orders of magnitude for T3 and five orders of magnitude for T4, showcasing the vast advantage of stochastic samplers compared to deterministic lattices for a high number of dimensions.

\subsubsection{Additional tests}

In this section we present two additional tests.

We have searched for the injections of all tests with a random template bank \cite{PhysRevD.79.104017} covering the follow-up region, with the number of likelihood evaluations (number of templates) twice as large as the value shown for \pt{} in figure \ref{fig:rescomT} for each test. We find that only a very small percentage of injections have converged (less than $5\%$), which means that the usage of MCMC or nested sampling methods is more efficient than a random template bank.

\begin{figure}[htbp]
  \centering
  \includegraphics[width=\columnwidth]{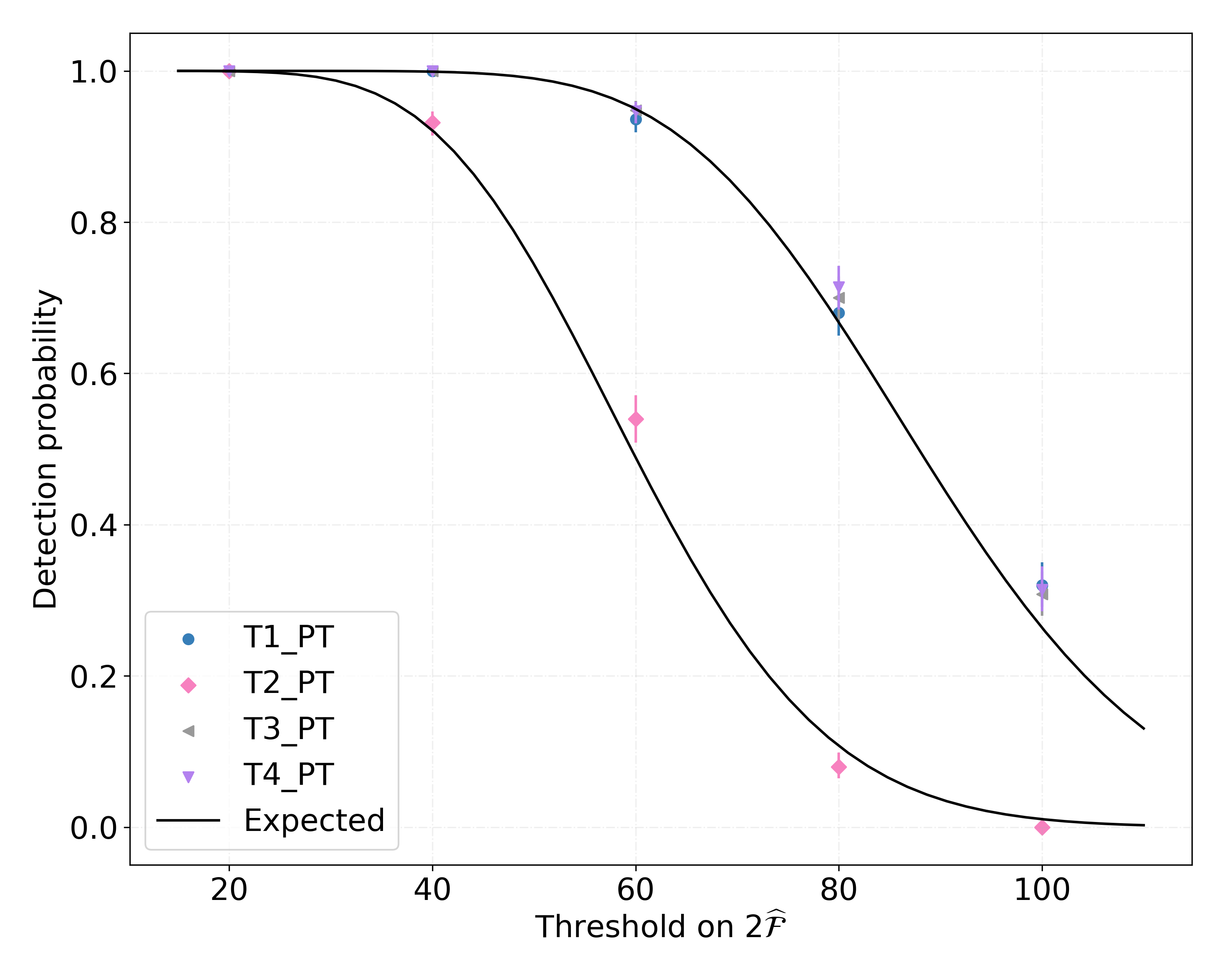}
  \caption{Detection probability as a function of the threshold on the $\Fsc$-statistic for the four different tests. The black lines show the theoretical expectation for the two different $\rho^2$ values of table \ref{tab:pars}. The markers show the results from the injections, with the vertical bars showing the 1$\sigma$ standard deviation.}
  \label{fig:rescomProbDet}
\end{figure}

As a final safety check, we compare the fraction of detected injections after setting five different thresholds on the $2\Fsc_{\mathrm{cand}}$ value of the candidate with the expected theoretical value. This expected value is calculated by using the survival function of the non-central $\chi^2$ distribution for the given thresholds, with the $\rho^2$ values of table \ref{tab:pars} as the non-centrality parameter. Figure \ref{fig:rescomProbDet} shows the result, both for the different tests and for the expected values. It can be seen that our results agree well with the theoretical expectation, indicating that the usage of the convergence criterion presented in section \ref{sec:convcrit} is safe.

\subsection{Box vs ellipse}
\label{sec:priorComparison}

The computational cost of a follow-up stage depends on the region $\mathcal{R}$ occupied by the prior distribution, as shown in the two previous sections. The region used in the previous tests and by the follow-up procedures of \cite{PhysRevLett.124.191102,PhysRevD.103.064017,Covas_2022} is given by a multi-dimensional box, where an independent distribution for each of the searched parameters is used as the prior.

In order to reduce the size of the follow-up region, we can instead use a multi-dimensional ellipsoid that takes into account the correlations between the different parameters given by the metric, as discussed in section \ref{sec:boxellipse}. 

\begin{figure}[htbp]
  \centering
  \includegraphics[width=\columnwidth]{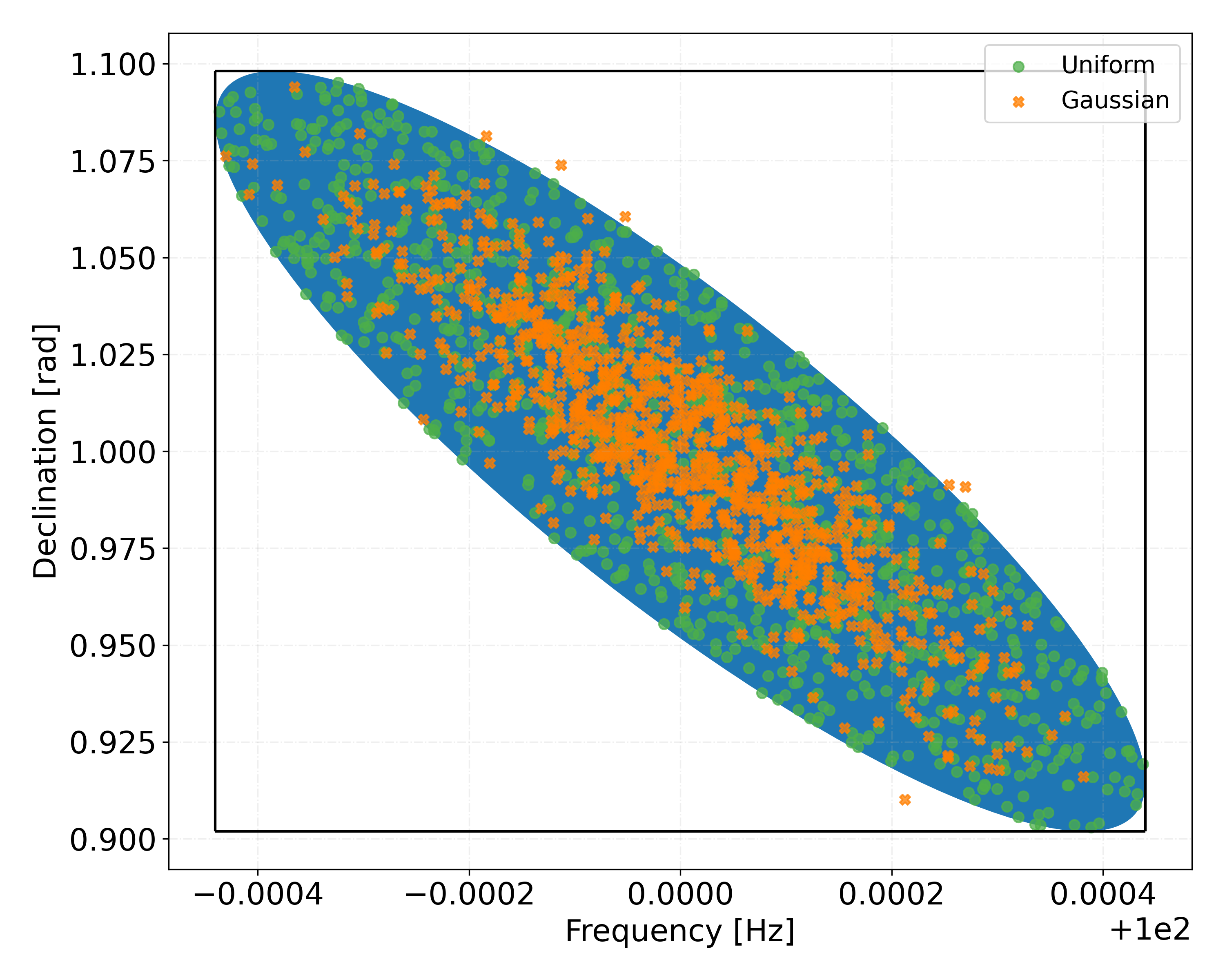}
  \caption{The blue region shows the ellipse with a maximum mismatch $m_{\mathcal{R}}=2$ for 180 days of data and $T\seg = 3600$ s, centered on $f_0 = 100$ Hz and $\delta = 1$ rad. The orange crosses show 1000 samples from the Gaussian distribution, with a coverage fraction of $x=0.99$ (6 out of 1000 samples are outside of the ellipsoid, with a maximum mismatch of 2.5). The green circles show 1000 samples from the uniform distribution, all contained inside the ellipsoid (with a maximum mismatch of 1.9998). The black lines mark the bounding box that encloses the ellipse. The number of unit-mismatch templates in the bounding box is $\mathcal{N}^*_{\mathrm{box}} = 17$, and $\mathcal{N}^*_{\mathrm{ell}} = 6$ for the ellipse. For a maximum mismatch between templates of $m_0 = 0.1$ and an $A^*_n$ lattice the number of templates is $\mathcal{N}_{\mathrm{box}} = 63$ and $\mathcal{N}_{\mathrm{ell}} = 24$.}
  \label{fig:EllipseGaussian}
\end{figure}

In the following tests we will use a uniform and a Gaussian distribution as our priors. For the Gaussian, the covariance $C_{ij}$ depends on the normalized mismatch metric $\bar{g}_{i j}$ (defined below) and on the desired coverage fraction $x$ (fraction of probability that is contained within the ellipsoid):
\begin{align}
  C_{i j} &= \frac{\bar{g}^{i j}}{c_f} = \frac{\bar{g}^{i j}}{2 \Gamma^{-1}(n/2, x)} \\
  \bar{g}_{ij} &\equiv \frac{ g_{ij} }{m_{\mathcal{R}}},
\end{align}
where $\bar{g}^{i j}$ is the inverse of the normalized metric, and $c_f$ is related to the coverage fraction $x$, given by the inverse of the cumulative distribution of the $\chi^2$ distribution with degrees of freedom equal to the number of searched parameters\footnote{For a Gaussian variable $x$, the contours of constant density are given by $(x-\mu)^T C_{i j} (x-\mu) = m$, which is a variable with a $\chi^2$ distribution.}, which is equal to the inverse of the regularized lower incomplete gamma function $\Gamma^{-1}$. Figure \ref{fig:EllipseGaussian} shows an example of samples generated from the uniform and Gaussian distributions and the corresponding ellipsoid with its bounding box.

In order to test the efficiency and safety of the uniform and Gaussian priors on the ellipsoid follow-up region, we perform the same sets of tests as in the previous section using the best \pt{} sampler configuration. The average number of unit-mismatch templates of the ellipsoid follow-up region is given in the second column of table \ref{tab:pars} (as previously explained, the number is different for each injection, because each used a different $m_{\mathcal{R}}$ in order to have a constant $\mathcal{N}^*_{\mathrm{box}}$). For these tests we shift the center of the prior from the signal parameters, so that the center of the of the Gaussian distribution is not artificially favored. To obtain the shifted center we draw a random sample within this ellipsoid using a uniform distribution\footnote{For the Gaussian distribution, this uniform drawing is not consistent with the usage of the Gaussian distribution itself, since a Gaussian would need to be used to draw the new center within the initial ellipsoid. We remark that the results obtained this way represent a more conservative assessment of the efficiency of the Gaussian distribution.}. This sample is the new center, while we keep the same ellipsoid size and shape.

\begin{figure}[htbp]
  \centering
  \includegraphics[width=\columnwidth]{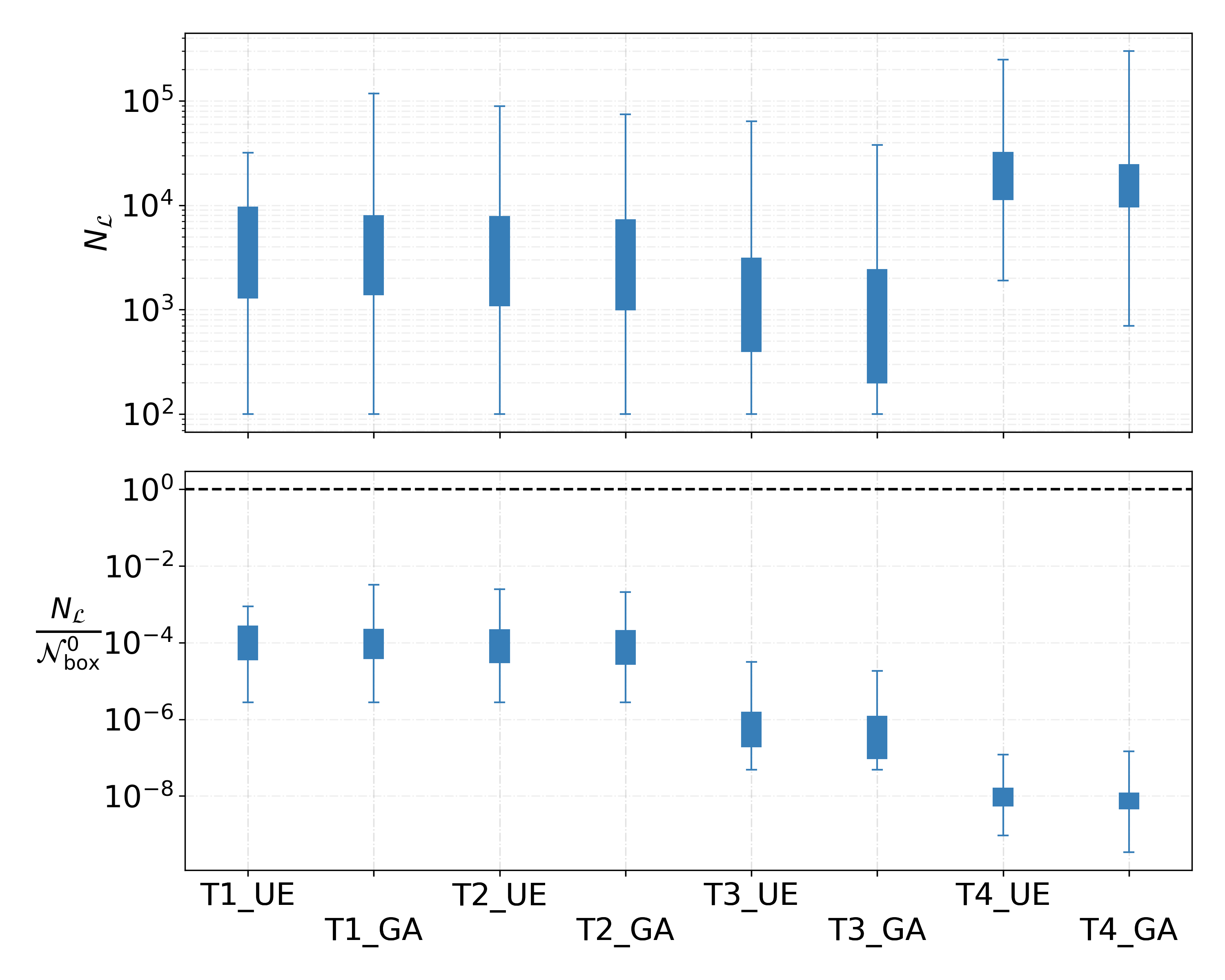}
  \caption{The upper plot shows the number of likelihood evaluations as a function of the test set and prior distribution (uniform ellipsoid and Gaussian with coverage fraction $x=0.99$), for the best \pt{} sampler configuration. The center of the prior has been displaced from the signal parameters. The labels on the x-axis refer to the different tests of table I and the different distributions, where UE refers to the uniform in the ellipsoid and GA to the Gaussian. The lower plot shows the same quantity but divided by $\mathcal{N}^{\,0}_{\mathrm{box}} \equiv \mathcal{N}(A^*_n,m_0=0.1)$, which is the number of templates required by a deterministic template bank with a maximum mismatch $m_0 = 0.1$ and an $A_n^*$ lattice. Each box extends from the 25 to the 75 quartiles, while the vertical lines go to the minimum and maximum of each distribution.}
  \label{fig:MultiPrior}
\end{figure}

The results are shown in figure \ref{fig:MultiPrior}. It can be seen that using a correlated prior with a smaller number of unit-mismatch templates reduces the required number of likelihood evaluations, in some cases up to three orders of magnitude. It can also be seen that using the Gaussian prior requires a similar number of likelihood evaluations than the uniform distribution.

\subsection{Computational efficiency}
\label{sec:comptests}

In this section we test the computational model explained in section \ref{sec:compmodel} and present the improvements in the computational efficiency of our framework.

\begin{figure}[htbp]
  \centering
  \includegraphics[width=\columnwidth]{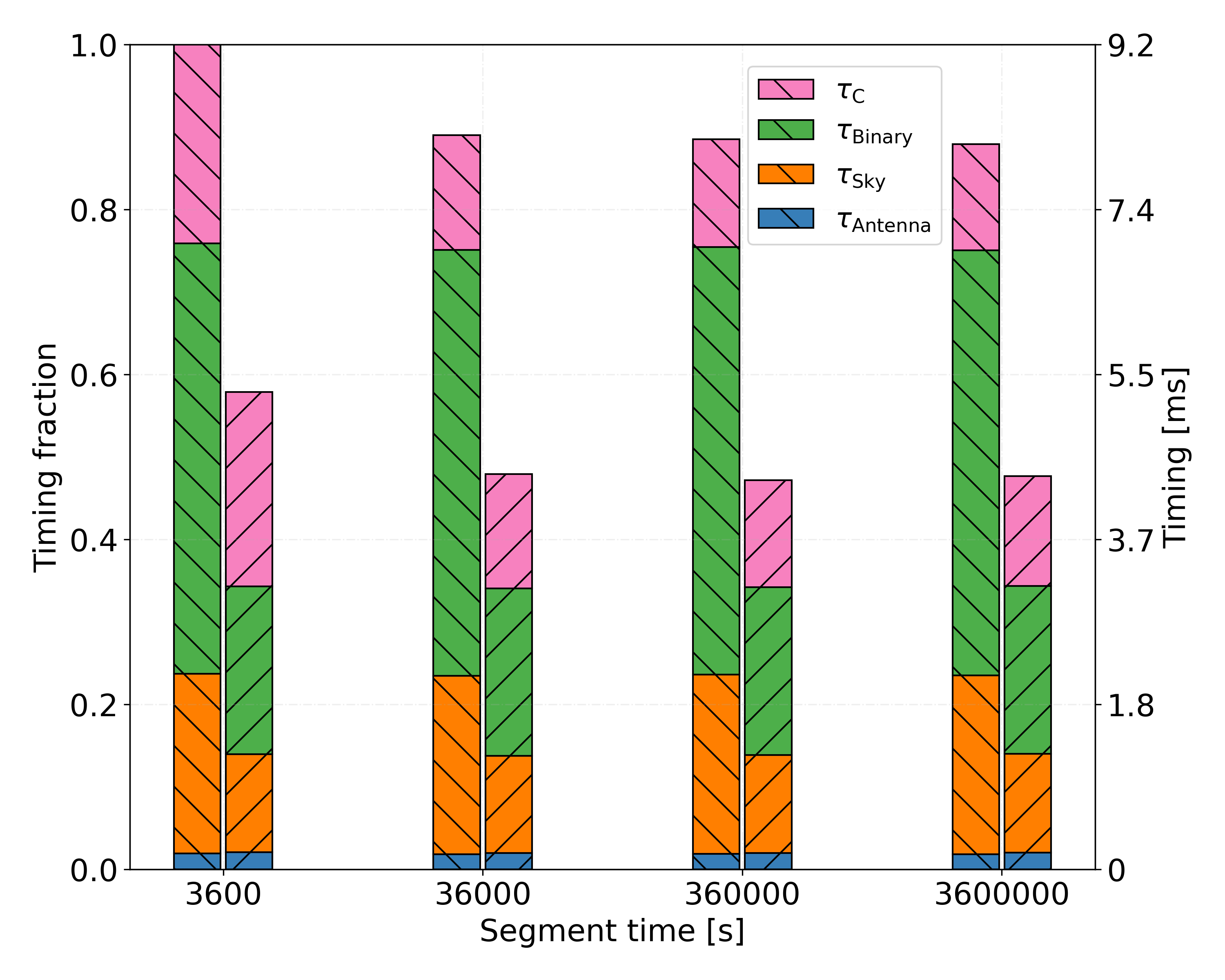}
  \caption{Fraction (respective to the maximum shown on the leftmost bar) and absolute time to compute the different components of $\tau_{\mathrm{F}}$, given by equation \eqref{eq:detstat}. For each of the 4 different segment times, the left bar shows the timings with the previous \ls{} code, while the right bar shows the timings with our optimized code. The four different colors in each bar show the different contributions in equations \eqref{eq:detstat} and \eqref{eq:buffer}, which from top to bottom are: $\tau_{\mathrm{C}}$, $\tau_{\mathrm{Binary}}$, $\tau_{\mathrm{Sky}}$, and $\tau_{\mathrm{Antenna}}$. The total amount of SFTs is $N_{\mathrm{SFTs}} = 20000$ with $T_{\mathrm{SFT}} = 1800$ s. These timings were obtained using an AMD EPYC 7351P 16-Core Processor.}
  \label{fig:TimingFraction}
\end{figure}

In a search with a lot of frequency or spin-down/up values, where the buffering quantities whose timings are given by equation \eqref{eq:buffer} can be reused multiple times, the main timing factor is usually the core time $\tau_{\mathrm{C}}$. In our follow-up case, where the next template is generated in a stochastic way, only one frequency and spin-down/up value is calculated at a time and the main timing factor is the buffer time $\tau_{\mathrm{B}}$. The left bars in figure \ref{fig:TimingFraction} show that the sum of the three buffering times that appear in equation \eqref{eq:buffer} amounts to a fraction between 0.6 and 0.7 of the total time. The most computationally expensive timing quantity is the source-to-BB calculation $\tau_{\mathrm{Binary}}$.

We have made several improvements to the code that calculates the semi-coherent detection statistic, mainly related to the computational efficiency of the barycentering calculations. The right bars in figure \ref{fig:TimingFraction} show the comparison to the previous code, where it can be seen that the improvement is almost a factor of two.

\begin{figure}[htbp]
  \centering
  \includegraphics[width=\columnwidth]{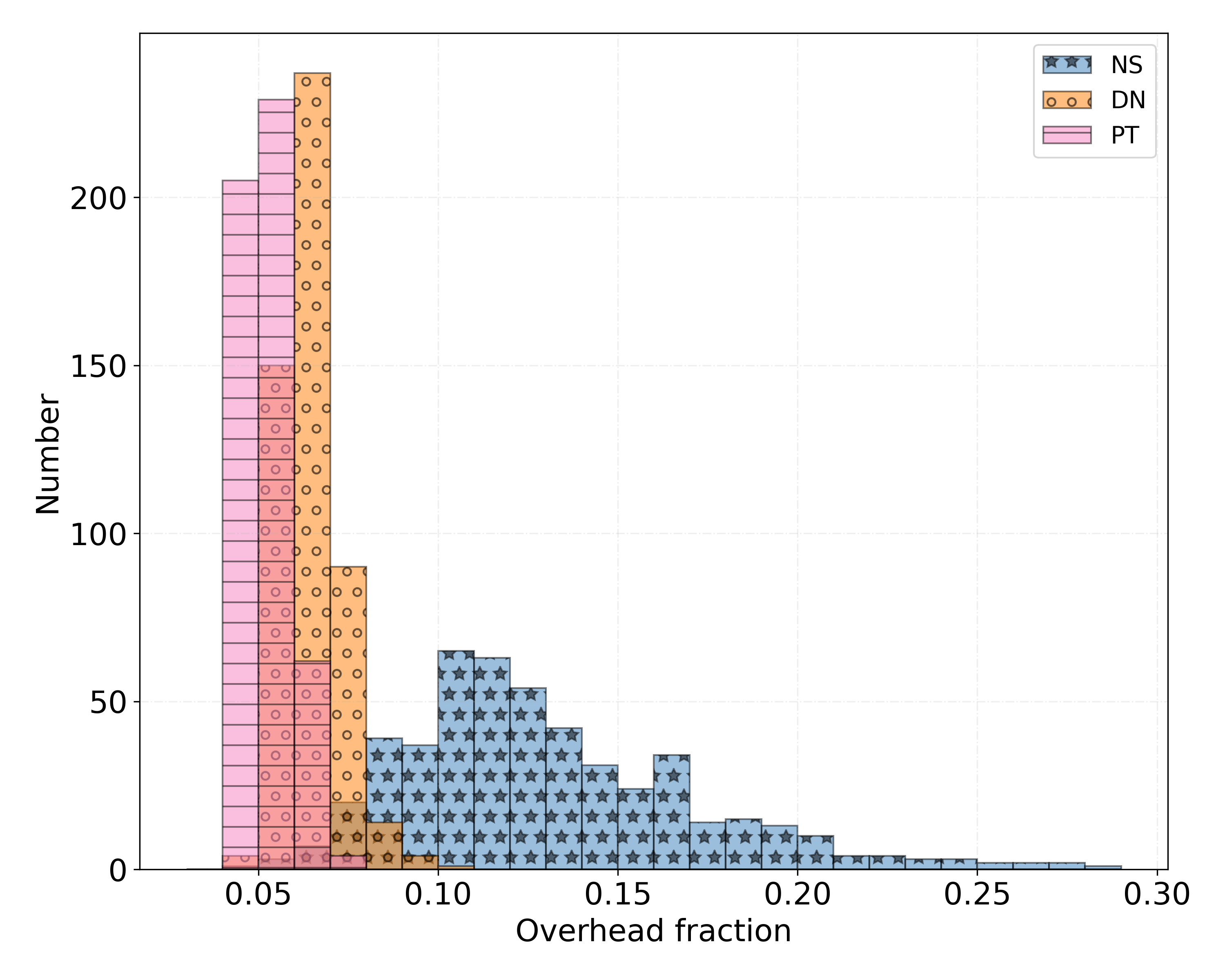}
  \caption{Histogram showing the fraction of time spent outside of the likelihood function in a given parameter-estimation run. The orange bars with circles show the results when using the \dn{} sampler, the blue bars with stars show the results using the \ns{} sampler, and the pink bars with lines show the results when using the \pt{} sampler (all using the results from the best configurations of section \ref{sec:comparing}. The total amount of SFTs for this test is $N_{\mathrm{SFTs}} = 17280$. These timings were obtained using an AMD EPYC 7351P 16-Core Processor.}
  \label{fig:Overhead}
\end{figure}

We also characterize the overhead time of each sampler, which is the time spent outside of the likelihood function. The results can be seen in figure \ref{fig:Overhead} (for $N_{\mathrm{SFTs}} = 17280$), which shows that the overhead fraction for the \pt{} and \dn{} samplers is between 0.03 and 0.1, while for the \ns{} sampler it is between 0.07 and 0.3. As explained in section \ref{sec:compmodel}, the cost of computing the likelihood depends linearly on the number of SFTs (for the $\mathcal{F}$-statistic method used in this paper). For this reason, the overhead fraction can vary greatly depending on the number of SFTs that are used, since for a higher number the $\tau_{\mathrm{F}}$ factor will increase while the overhead factor $\tau_{\mathrm{O}}$ will be constant. For example, for a duration of 10 days with 1800 s SFTs ($N_{\mathrm{SFTs}} = 480$), we observe instead that the overhead fraction increases up to $\sim 0.7$ from the previous $\sim 0.1$ value.

\begin{figure}[htbp]
  \centering
  \includegraphics[width=\columnwidth]{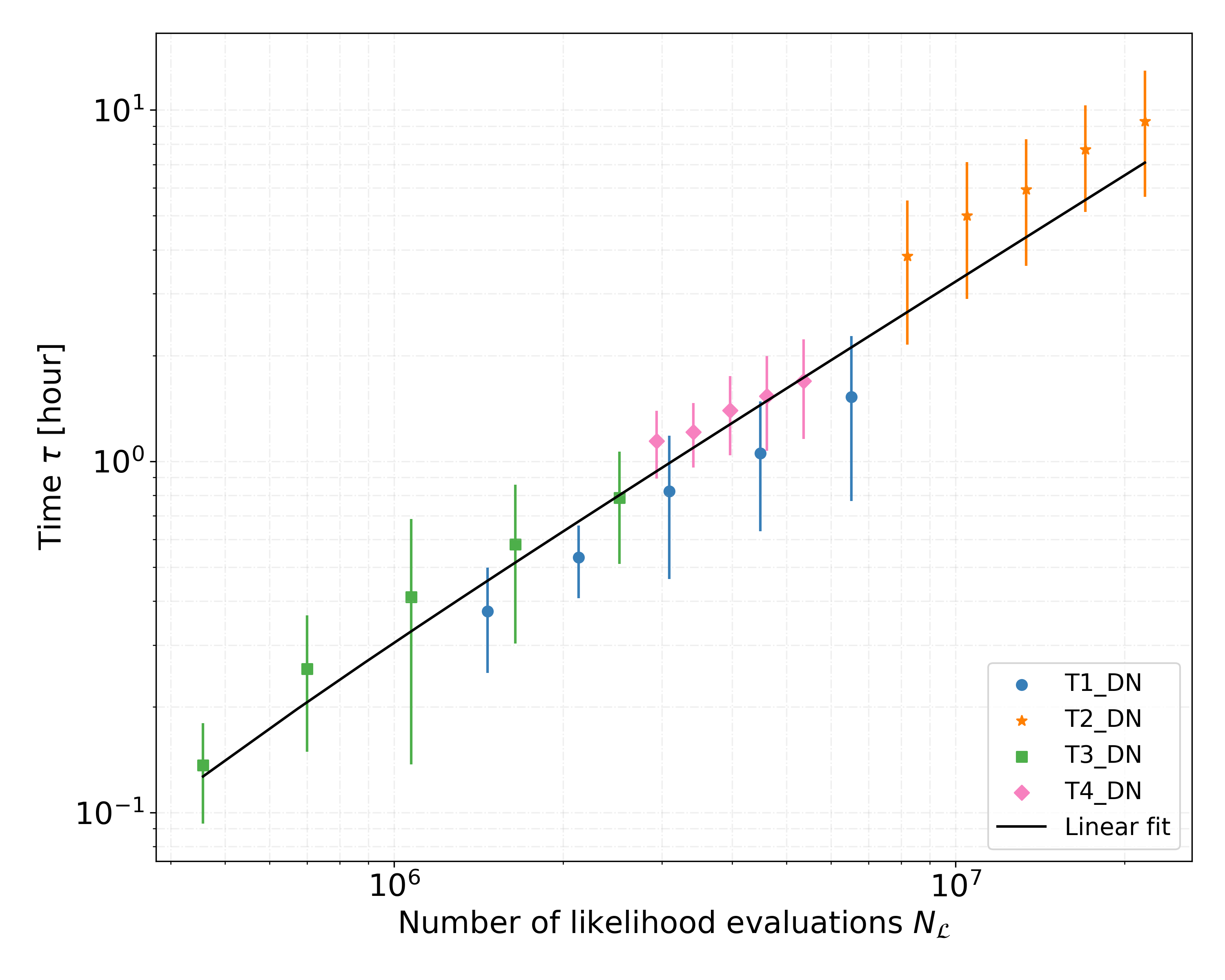}
  \caption{Total runtime given by equation \eqref{eq:tau} as a function of the number of likelihood evaluations for the \dn{} sampler. The different markers represent the different test sets from section \ref{sec:comparing}, while the black line shows the results of a linear fit, with a slope of $\bar{\tau} = 3.27 \times 10^{-7}$ hours. For each test set, we have created ten different $N_{\mathcal{L}}$ bins (each marker shows the bin mid-point), and we have calculated the mean run-time for the injections belonging to each bin. The vertical bars show the 2$\sigma$ standard deviation of each bin. These timings were obtained using an AMD EPYC 7452 32-Core Processor.}
  \label{fig:NlikeTime}
\end{figure}

In figure \ref{fig:NlikeTime} we show the total runtime given by equation \eqref{eq:tau} as a function of the number of likelihood evaluations for the four different test sets and \dn{}, with a linear fit to these results. We can see that a linear relationship between these two quantities is followed, with a mean timing per likelihood evaluation of $\bar{\tau} = 3.27 \times 10^{-7}$ hours. The vertical spread can be explained by three different factors: the different times $\tau_{F}$ that are needed to calculate the likelihood, depending on the region of parameter space; the different overhead of the sampler, which depends on the number of dimensions; the intrinsic spread due to the varying activity in the supercomputer that was used to carry out the runs.

Although the samplers that we have used allow for multi-core parallelization, we have not tested this feature. In typical follow-up scenarios we may have to analyze on the order of millions of candidates, so parallelizing each individual parameter-estimation run might not be a useful strategy.

\section{Conclusions}
\label{sec:conclusions}

In this paper we have presented a new framework to follow-up candidates from CW searches. This framework expands the capabilities of \pf{} in a number of ways: more flexibility in the choice of sampler and prior distribution, and a new convergence criterion. We have shown that for a large number of dimensions we are able to perform searches of CWs with a greatly reduced computational cost as compared to a search that would use a template bank. Furthermore, we have shown that it is possible to find the maximum posterior point for parameter-space regions much larger than previously thought. We have focused on finding a good configuration of the \pt{}, \dn{}, and \ns{} samplers in order to reduce the computational cost of a single follow-up stage, showing that they can produce similar results with comparable computational efficiency. We have also characterized the computational cost of a parameter-estimation run, and shown the improved computational efficiency of our framework.

The main drawback of using stochastic samplers, as compared to a follow-up with a deterministic template bank, is that the samplers need to be characterized (for example the number of live points or the number of walkers) since the number of required likelihood evaluations depends on the size of parameter space and the strength of the signals. 

There are many ways in which this work could be expanded. We have tested a limited number of samplers and sampler configurations, so other options could be more efficient than the ones we have tried. Furthermore, usage of proposals (to find the next sampled point) specific to the CW problem (in a similar way to the compact binary merger proposals explained in \cite{bilby_mcmc}) could increase the efficiency of the samplers. Another option would be to use block-updating proposals, where only a subset of parameters are updated at each step. This could help in decreasing the computational cost, for example by keeping the sky position and binary parameters fixed and not recomputing the barycentering transformations and the antenna-pattern coefficients for a fixed number of steps. Moreover, a different set of coordinates could improve the sampling efficiency, such as using the sky coordinates proposed in \cite{Wette_2015} instead of the equatorial coordinates $\alpha$ and $\delta$. Finally, these stochastic sampling algorithms could be ported to GPUs to further improve their computational efficiency.

The main goal of MCMC and nested sampling algorithms is not to find the maximum posterior point. It would be interesting to compare the efficiency of these algorithms against other tools specifically designed to locate maxima and minima, such as \textit{NOMAD} \cite{Shaltev_Prix_2013}.

Besides classical MCMC and nested sampling algorithms, machine learning is becoming widely used for gravitational-wave searches and parameter estimation, showing large decreases in the computational cost. This could also be tried with CWs, by using an algorithm similar to \textit{DINGO} \cite{Dax_2021}. 

Lastly, this framework could be used to search for spin-wandering of sources in binary systems. A reversible jump MCMC \cite{PhysRevD.101.123021} that traverses over different models with different numbers of parameters could be proposed to measure this effect, or in combination with a Kalman filter, as recently done in \cite{Melatos_2023}.

\begin{acknowledgments}
This project has received funding from the European Union's Horizon 2020 research and innovation programme under the Marie Sklodowska-Curie grant agreement number 101029058.

This work has utilized the ATLAS computing cluster at the MPI for Gravitational Physics Hannover.
\end{acknowledgments}

\bibliography{Paper}

\appendix
\section{Illustration of final-stage parameter-estimation}
\label{sec:hardware}

In this appendix we illustrate that our framework is also capable of obtaining precise posterior distributions with real data, a procedure that might be used at the last stage of a follow-up. To achieve this, we show the results of a parameter-estimation run in a region around one of the hardware injections with binary parameters in Advanced LIGO O3 data \cite{Biwer_2017}, using both the \dn{} and \pt{} samplers with the default tuning parameters in \bb{} 2.0.3. The parameter-space region that we analyze is the bounding box (centered on the hardware injection parameters) of the mismatch ellipse (given by equation \eqref{eq:phasemetric}) with $m_{\mathcal{R}} = 3$, with uniform priors for each of the six searched parameters. Here we use the default termination criteria for each sampler: for \dn{}, the estimated remaining log-evidence needs to be smaller than 0.1; for \pt{}, 5000 independent posterior samples are required.

\begin{figure*}[htbp]
  \centering
  \includegraphics[width=\textwidth]{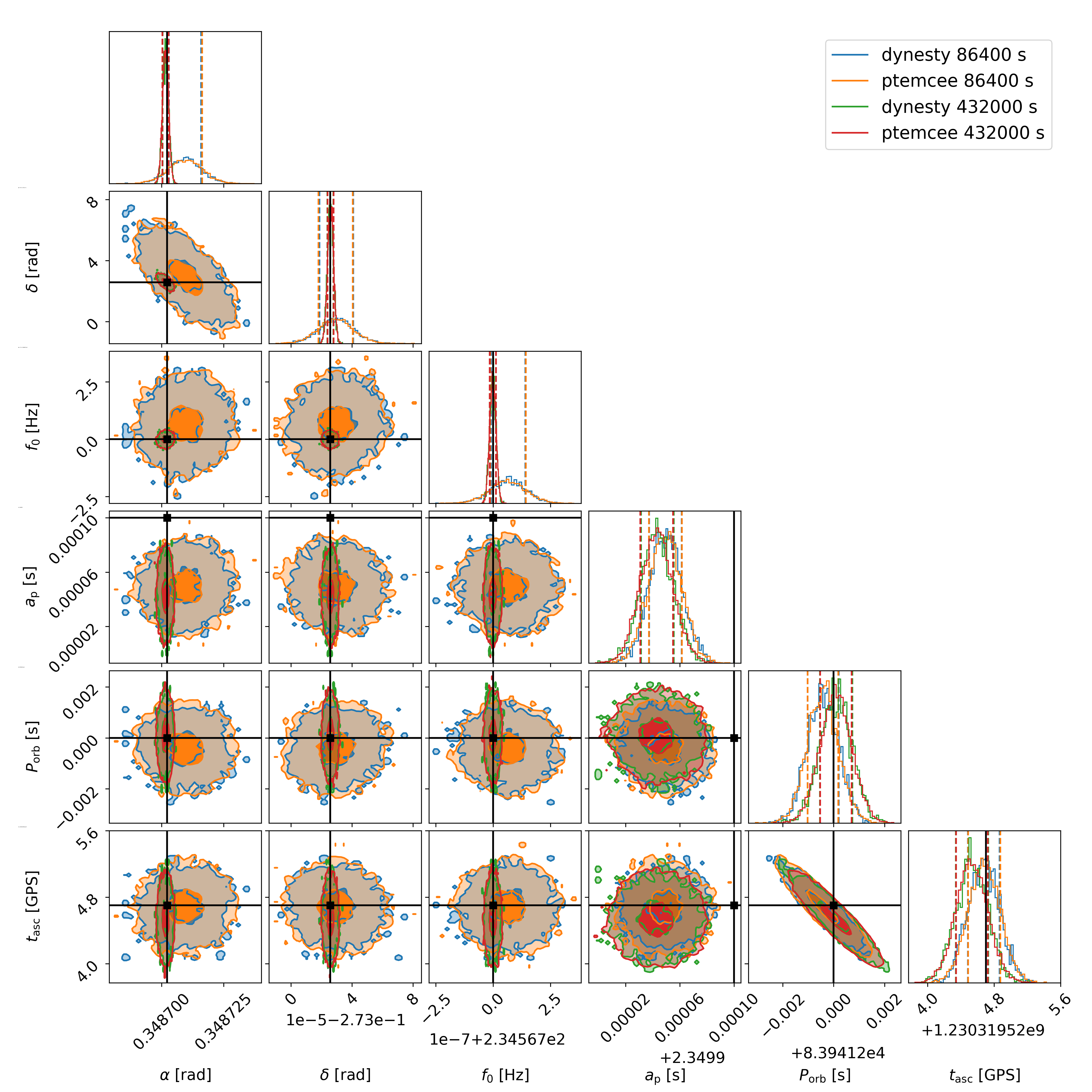}
  \caption{Corner plot for the hardware injection\footnote{See O3 Continuous Wave Hardware Injections section in \url{https://www.gw-openscience.org/O3/o3_inj}} number 16 with binary parameters using Advanced LIGO O3 data with SFTs of $T_{\mathrm{SFT}} = 200$ s. We have used the \dn{} and \pt{} samplers with their default parameters in \bb{} 2.0.3. The black lines and points show the parameters of the signal. The other four different colors show the posterior distributions for the two samplers, for two coherent times of 86400 and 432000 seconds.}
  \label{fig:HI0}
\end{figure*}

The results can be seen in figure \ref{fig:HI0}, where it is shown that the two samplers produce comparable results, and are able to recover the posterior of the signal. We show the results from two runs with different coherent times, $T\seg = 86400$ s and $T\seg=432000$ s, respectively. It can be seen that when the coherent time is increased, the uncertainty is reduced in some of the parameters, following the expected behaviour derived from the parameter-space metric given by equation \eqref{eq:metric} \cite{2015arXiv150200914L}.

\end{document}